\documentclass[bibyear]{aa} 

\usepackage{graphicx}
\usepackage{txfonts}
\usepackage{cite}
\usepackage{amsmath}	
\usepackage{amssymb}	
\usepackage{epsfig}
\usepackage{color}
\usepackage{longtable,lscape}
\usepackage{natbib}
\bibpunct{(}{)}{;}{a}{}{,} 
\usepackage{graphicx}	
\usepackage{enumitem}
\usepackage[colorlinks=true, linkcolor=blue, citecolor=blue, filecolor=blue, urlcolor=blue]{hyperref}
\newcommand{\HI}{\mbox{H\,{\sc i}}}

\newcommand{\J}{$J$}
\newcommand{\HH}{$H$}
\newcommand{\K}{$K_{\rm s}$}
\newcommand{\Ko}{$K^{\rm o}_{\rm s}$}
\newcommand{\ko}{$K^{\rm o}_{\rm s}$}

\newcommand{\Kd}{$K^{\rm o,d}_{\rm s}$}
\newcommand{\kd}{$K^{\rm o,d}_{\rm s}$}

\newcommand{\ebv}{$E(B-V)$}
\newcommand{\ak}{A_{\rm K}}
\newcommand{\ab}{A_{\rm B}}
\newcommand{\hk}{$(H-K_{\rm s})$}
\newcommand{\jk}{$(J-K_{\rm s})$}

\newcommand{\hko}{$(H-K_{\rm s})^{\rm o}$}
\newcommand{\jko}{$(J-K_{\rm s})^{\rm o}$}

\newcommand{\hkoc}{$(H-K_{\rm s})^{\rm o,c}$}
\newcommand{\jkoc}{$(J-K_{\rm s})^{\rm o,c}$}

\newcommand{\kms}{\mbox{km\,s$^{-1}$}}

\def\approxlt{\lower.2em\hbox{$\buildrel < \over \sim$}}  
\def\approxgt{\lower.2em\hbox{$\buildrel > \over \sim$}}  

\newcommand{\nan}{Nan\c{c}ay}

\newcommand{\fa}{{$f_{\rm a}$}}

\newcommand{\cf}{{cf.}\ }
\newcommand{\eg}{{e.g.},\ }

\definecolor{grey}{rgb}{0.5,0.6,0.7}

\begin{document}

\title{Supplementing the 2MRS survey: The 2MZoA catalogue}

\author{A.C. Schr\"oder \inst{\ref{inst1}}
          }

\institute{Max-Planck-Institut f\"ur extraterrestrische Physik, Gie{\ss}enbachstra{\ss}e 1, 85748 Garching, Germany, \email{anja@mpe.mpg.de} \label{inst1}
             }

   \date{Received ...; accepted ...}

  \abstract
   {
We present a homogeneous catalogue of verified 2MASX galaxies in high
extinction areas, which is complete to a Galactic extinction-corrected
magnitude of \Ko\ $\le 11\fm75$. It covers the low Galactic latitudes ($|b|
\le 10\fdg0$), called Zone of Avoidance (ZoA), and areas of high foreground
extinctions (\ebv $> 0\fm95$) at higher latitudes. This catalogue
supersedes the previously presented bright 2MZoA catalogue which was only
complete to \Ko\ $\le 11\fm25$. It fully complements the 2MASS Redshift
Survey (2MRS) galaxy catalogue, which has the same magnitude limit but
excludes high extinction regions. The combination of the two catalogues,
the extended 2MRS or e2MRS, is a uniquely whole sky redshift survey of
galaxies and forms a sound basis for studies of large-scale structures,
flow fields, and extinction across the ZoA.

The here presented catalogue comprises 6899 galaxies with 6757 galaxies at
low latitudes and 142 galaxies in highly obscured high-latitude areas.  The
completion rate in redshifts is almost 75\%. The catalogue is complete up
to star density levels of at least $\log N_*/{\rm deg}^2 < 4.3$, but the
completion rate of the fainter part is affected by foreground extinction at
all levels. This can be rectified by using a diameter-dependent extinction
correction, adding 605 highly obscured but apparently faint galaxies (with
\Ko\,$ > 11\fm75$ and \Kd\,$ \le 11\fm75$) to the sample. This extended
sample shows good completion rates with extinction up to at least $\ak <
1\fm3$. Omission of such a diameter-dependent extinction correction may
lead to a biased flow field even at intermediate extinction values as found
in the 2MRS survey. As in our previous investigations, we recommend a
correction factor $f=0.86$ to be applied to the extinction maps, and we
find it to be independent on Galactic longitudes or latitudes.
}


\maketitle

\nolinenumbers

\section{Introduction}   \label{sec:introduction}  

Our knowledge of the local Universe, that is, the local large scale
structures and related dynamics, is still incomplete due to the
difficulties in finding and studying galaxies near the Galactic plane, the
so-called Zone of Avoidance (ZoA). Dust extinction as well as star crowding
severely hinder both the identification of galaxies (where different
wavelengths are affected by different biases, \eg \citealt{kraan00} and
references therein) as well as follow-up observations to obtain, for
example, redshifts \citep[e.g.,][]{macri19}.

The ZoA is known to obscure and bisect major parts of dynamically important
structures such as the Perseus-Pisces Supercluster
\citep[][]{focardi84,chamaraux90}, the Great Attractor
\citep[][]{dressler87,kraan96,woudt04}, and the Local Void
\citep[][]{tully08,tully19}. Even to date, newly identified low-latitude
structures are still being discovered, for example the extended Vela
supercluster \citep[][]{kraan17} at $cz \sim 18,500\,\kms$, contributing to
the local bulk flow motion \citep[e.g.,][]{hudson04,springob16}.

While \HI\ surveys are best in penetrating the foreground dust, they mainly
trace filaments through gas-rich spiral galaxies. In contrast, NIR surveys
serve better in getting a more comprehensive view of the local large-scale
structures. Foreground extinction at these wavelengths is a lesser problem,
and instead the detection of galaxies is more likely constrained by star
crowding \citep[e.g.,][hereafter Paper I]{schroeder19}.  The only truly
all-sky NIR survey is the 2MASS (Two Micron All Sky Survey;
\citealt{skrutskie06}) with a catalogue of extended sources, 2MASX
\citep{jarrett04}. A subset of this catalogue, restricted by Galactic
latitude and foreground extinction, led to the 2MASS Redshift Survey (2MRS,
\citealt{huchra12}).

2MASS is a comprehensive though shallow survey, but deeper NIR surveys,
covering only the ZoA, exist, \eg the UKIDSS GPS (Galactic Plane Survey;
\citealt{lucas08}), the Vista VVV (VISTA Variables in the V\'ia Lactea;
\citealt{minniti10}) as well as VVVX (the extended VVV; \citealt{saito24}),
and the DECam Plane Survey \citep{schlafly18}. With their higher resolution
they are ideal for detecting galaxies, but by-eye searches or even
verification of automatically detected extended sources of such a data
volume is not feasible anymore and one needs to rely on other methods, like
machine learning \citep[e.g.,][]{daza23, alonso25}.

To produce the extended 2MRS, e2MRS, which is the first nearly whole-sky
redshift survey, our group has started to complement 2MRS by identifying
2MASX galaxies deep in the ZoA and measuring redshifts for them. As a pilot
study, Paper I presents the 2MZoA catalogue of verified 2MASX galaxies down
to an extinction-corrected magnitude limit of \ko\ $=11\fm25$, which is
complete up to star density levels of $\log N_*/{\rm deg}^2 <4.5$, thus
reducing the ZoA to $2.5-4$\% of the whole sky. \citet{kraan18} observed
useful candidates with the \nan\ Radio Telescope (NRT), resulting in 232
detections. An optical spectrographic survey is still
ongoing. \citet[][hereafter Paper II]{schroeder21} has used this sample to
investigate the Galactic foreground extinction and recommend the extinction
map presented by \citet{planck16b} with a correction factor $f =
0.86\pm0.01$. In this paper, we present the extension of the 2MZoA
catalogue down to an extinction-corrected magnitude limit of
\ko\ $=11\fm75$, so as to be fully complementary to 2MRS.  Obtaining the
missing redshifts will be more difficult, especially since many of the
faintest galaxies are too obscured for optical spectroscopy and will
require (N)IR spectroscopy or \HI\ observations.

This paper is organised as follows. In Sect.~\ref{sample} we describe the
sample, in Sect.~\ref{cat} we present the catalogue and in Sect.~\ref{prop}
we investigate its properties. Section~\ref{comp} discusses differences and
commonalities with Paper I and of various subsamples. In Sect.~\ref{galext}
we revisit the investigation on Galactic foreground extinction from Paper
II, using the new sample. Sect.~\ref{summary} finishes with a
summary. Online tables and some plots are presented in the appendix.

\section{The sample }  \label{sample}  

Details of the sample selection and extraction can be found in Paper I. For
convenience, we give here a short summary and point out the differences due
to the new selection criteria.

Our new magnitude limit is \Ko\,$ \le 11\fm75$. Contrary to the 2MRS sample
and our bright 2MZoA sample, we now choose to use a more suitable
extinction map to correct for Galactic extinction, as investigated in Paper
II. Instead of using the more common \citet{schlegel98} maps (hereafter
SFD), we recommended the less biased \citet{planck16b} maps, which were
reduced using the generalised needlet internal linear combination (GNILC)
component-separation method, hereafter called GN extinction.  Furthermore,
we apply a correction factor $f =0.86$ to the \ebv\ values as also
recommended in Paper II. For the 2MRS sample, the difference between the
SFD and GN extinction corrections is considered negligible since they
restrict their sample to areas where \ebv\,$ < 1\fm0$. The difference to
the bright 2MZoA sample will be more noticeable, with new galaxies being
included in the sample and others excluded.

In addition, with more redshift information and deep NIR images that have
become available since Paper I, some of the object classifications have
changed.  We will thus give a fully updated catalogue of all galaxies with
\Ko\,$ \le 11\fm75$ and include flags to indicate whether a galaxy is in
the old and/or the new catalogue.

As in Paper I, we have compiled two subsamples:
\begin{description}
\item [ZOA sample:] galaxies at $|b| \le 10\fdg0$ at all Galactic
  longitudes;

\item [EBV sample:] all objects at $|b| > 10\fdg0$ with \ebv $\ >
  0\fm95$. The limit refers to SFD extinctions as to be as close to the
  2MRS selection as possible; the overlap range has been chosen to cover
  any small variations in the actual calculation of \ebv.
\end{description}

\subsection{Sample extraction}

\begin{table}
\caption[]{Object samples at different stages during extraction.
\label{tabsamples}}
{\scriptsize
\begin{tabular}{lrrrr}
\hline
\noalign{\smallskip}
Sample    &   \multicolumn{2}{c}{2MZoA}  & \multicolumn{2}{c}{EBV} \\
          &   Paper I  &   this paper    & Paper I & this paper \\
\noalign{\smallskip}
\hline
\noalign{\smallskip}
2MASX extraction           & 146\,174  &  (same)  & 4248 & (same) \\
all objects, bright sample     &   6913    &   6080   &  502 &   424  \\    
all objects, faint sample      &   --      &   3859   &  --  &   130  \\    
galaxies, bright sample    &   3671    &   3470   &  88  &    76  \\    
galaxy candidates, bright sample &  4  &     3    &   0  &     0  \\    
galaxies, faint sample     &   --      &   3287   &  --  &    66  \\    
galaxy candidates, faint sample  & --  &     6    &  --  &     0  \\    
galaxies total             &   3671    &   6757   &  88  &   142  \\    
\noalign{\smallskip}
\hline
\end{tabular}
}
\end{table}

From the online 2MASX catalogue\footnote{See the 2MASS All-Sky Extended
  Source Catalog (XSC) as found online at
  http://irsa.ipac.caltech.edu/cgi-bin/Gator/nph-dd}, we extracted all
sources with the above mentioned selection criteria as well as the default
SQL constraints given in the webform. The resulting sample is the same as
in Paper I, but we now apply the GN extinction correction to extract the
magnitude-limited sample of objects, using the following extinction
conversion factors \citep{fitz99}: $A_\lambda$ / $\ab = 0.208, 0.128,
0.087$ for \J , \HH\ and \K -band, repetitively. Table~\ref{tabsamples}
lists the numbers of objects in the respective samples ZOA and EBV for the
original catalogue (Paper I) and for the present catalogue. For better
comparison we also give numbers for the bright sample (\Ko\,$ \le 11\fm25$)
and the faint sample ($11\fm25 < K^{\rm o}_{\rm s} \le 11\fm75$)
separately.

All 2MASX catalogue entries obeying the selection criteria have been
inspected visually on optical and NIR images and given a flag according to
the confidence we have that they are galaxies. Compared with Paper I, the
availability of deep NIR survey images has increased\footnote{\eg UHS
  (UKIRT Hemisphere Survey, \citealt{dye18}) and VVVX \citep{saito24}
  survey images have become available in the meantime.}, with 89\% of the
ZOA objects and 76\% of the EBV objects now covered.

As in Paper I, we extracted redshift information from online data bases and
literature, which also helped with the classification (see Appendix A for
details). Finally, we visually estimated galaxy types regarding the
presence or absence of a disc (mainly in preparation for \HI\ follow-up
observation); the B- and R-band images proved most useful for
low-extinction galaxies, while the deep K-band images were most helpful in
high-extinction areas. In addition to Paper I, we flag known and suspected
active galaxies (AGNs) since these often have a redder colour. We have also
revisited the extinction flag and reevaluated those for the bright sample.

As a result, we have a total of 6899 galaxies, with 6757 in the ZOA sample
and 142 in the EBV sample. In addition, there are nine galaxy candidates,
all in the ZOA sample. The differences to Paper I in the bright sample
source numbers are mainly due to the here applied extinction correction
factor (which reduces the extinction correction, resulting in fainter
corrected magnitudes) and, to a smaller degree, due to differences between
the SFD and GN extinction values.

\subsection{Subsamples and supplementary samples}

In the catalogue, we give flags for various sample definitions for
convenience. As already indicated in Table~\ref{tabsamples}, we can define
a bright and a full sample (with \Ko\,$ \le 11\fm25$ and \Ko\,$ \le
11\fm75$, respectively). For easier comparison with Paper I, we give an
`original sample' flag based on SFD extinction correction with no
correction factor (note that the number of galaxies in this sample still
differs from Paper I due to changes in galaxy class settings). We also give
a flag for the full sample but with no correction factor applied to the
\ebv\ values, which makes it easier to investigate this factor further if
there is a need.

\begin{table*}[ht]
\caption[]{Various sample flags 
\label{tabflags}}
{\scriptsize
\begin{tabular}{llcrrrrrrr}
\hline
\noalign{\smallskip}
Sample  & Selection & Flag & \multicolumn{3}{c}{ZOA} & \multicolumn{2}{c}{EBV} &
\multicolumn{2}{c}{Total}\\
 & & & Objects & Galaxies & Cand. & Objects & Galaxies & Objects & Galaxies \\
\noalign{\smallskip}
\hline
\noalign{\smallskip}
Full sample S                      & \Ko\,$\le 11\fm75$, GN, $f=0.86$ & f$_{\rm sam}$  & 9939 & 6757 &  9 & 554 & 142 & 10493 & 6899 \\ 
Bright (original) sample, Paper I  & \Ko\,$\le 11\fm25$, SFD, $f=1.0$ & f$_{\rm orig}$ & 6913 & 3693 &  3 & 502 &  89 &  7415 & 3782 \\ 
Bright sample, this paper          & \Ko\,$\le 11\fm25$, GN, $f=0.86$ & f$_{\rm br}$   & 6080 & 3470 &  3 & 424 &  76 &  6504 & 3546 \\ 
Full sample, $f=1.0$               & \Ko\,$\le 11\fm75$, GN, $f=1.0$  & f$_{\rm no-f}$ &10447 & 7099 & 14 & 585 & 152 & 11032 & 7251 \\ 
\noalign{\smallskip}
\multicolumn{9}{l}{With diameter-dependent extinction corrections:}\\
Extended sample S$e$              & \Kd\,$\le 11\fm75$, GN, $f=0.86$ & f$_{\rm ext}$ & 10960 & 7344 & 16 & 616 & 160 & 11576 & 7504 \\ 
Bright extended sample, Paper I & \Kd\,$\le 11\fm25$, SFD, $f=0.83$&f$_{\rm ext-orig}$& 7158 & 3734 &  5 & 515 &  90 &  7673 & 3824 \\ 
\noalign{\smallskip}
\hline
\end{tabular}
}
\end{table*}

We list the sample flags with their definitions as well as the numbers of
objects and of galaxies in Table~\ref{tabflags}. Note that the sample flags
do not distinguish between the ZOA and EBV sample since these can be simply
selected by Galactic latitude.

As mentioned in Paper I, the isophotal diameter of a galaxy is also
affected by extinction, and an additional, diameter-dependent, extinction
correction should be applied. This correction, however, has large
uncertainties and is best applied to the individual galaxy photometry by
extrapolating the radial brightness curves \citep{riad10}. We therefore do
not use it in our recommended sample. Instead, we have compiled an extended
sample based on these corrections, which will help to understand the
effects of applying such a correction on sample size and characteristics.
We give two such sample flags: one is based on the faint limit \Kd\,$ \le
11\fm75$, with the GN extinction correction and a factor $0.86$ applied,
the other refers to the sample presented in Paper I and is based on the SFD
extinctions (\Kd\,$ \le 11\fm25$), with a correction factor $f=0.83$
applied.

\section{The catalogue } \label{cat}  

The catalogue with a total of 10960 ZOA and 616 EBV objects is available
online at CDS; a few example rows are listed in Table~\ref{tabzoa} in the
appendix. It supersedes the catalogues given in Paper I. The table comes in
two sections: (a) full sample as per our selection criteria, and (b)
supplementary sample for objects that obey the limit \Kd\,$ \le 11\fm75$
but not \Ko\,$ \le 11\fm75$ when a diameter-dependent extinction correction
is applied\footnote{A third section comprises 12 additional galaxies
  neither included in the full nor in the supplementary sample but which
  have \Ko\,$ \le 11\fm75$ if no correction factor is applied to the
  extinction; we only list these for the comparison of samples, they are
  not used elsewhere.}. Columns are the same as in Paper I except for a
different suite of sample flags, listed in Col.~7, and an updated object
flag in Col 6, which now includes an indicator for AGNs. We have also added
the best velocity measurement found in the literature, and include detailed
flags separately for optical and \HI\ velocities in Col.~10.

The columns are as follows:
\begin{description}

\item{Col.\ 1:} ID: 2MASX catalogue identification number (based on J2000.0
  coordinates).

\item{Col.\ 2a and 2b:} Galactic coordinates: longitude $l$ and latitude
  $b$, in degrees.

\item{Col.\ 3:} Extinction: \ebv\ value derived from the GNILC Planck maps
  \citep{planck16b}, in magnitudes.

\item{Col.\ 4:} Object class: 1 = obvious galaxy, 2 = galaxy, 3 = probable
  galaxy, 4 = possible galaxy, 5 = unknown, 6 = lower likelihood for
  galaxy, 7 = unlikely galaxy, 8 = no galaxy, 9 = obviously not a galaxy.

\item{Col.\ 5:} Object offset flag: `o' stands for coordinates that are
  offset from the centre of the object, and `e' stands for a detection near
  the edge of an image (at an offset position from the object centre; the
  properly centred object is detected on the adjacent image).

\item{Col.\ 6:} Object flag: `p' stands for Planetary Nebula (PN), `q'
  indicates a QSO/AGN, and `b' stands for blazar; tentative classifications
  are indicated by a question mark.

\item{Col.\ 7:} Sample flags in the following order:
\begin{enumerate}[label=(\alph*)]
\item f$_{\rm gal}$: Galaxy flag: `g' denotes a galaxy (object classes
  $1-4$), and `p' stands for galaxy candidates (class 5);
\item f$_{\rm sam}$: Full sample (\Ko\,$ \le 11\fm75$, GN extinction
  correction, $f=0.86$);
\item f$_{\rm orig}$: Bright sample as in Paper I (\Ko\,$ \le 11\fm25$, SFD
  extinction correction, no correction factor);
\item f$_{\rm br}$: Bright sample, this paper, for comparison with Paper I
  (\Ko\,$ \le 11\fm25$, GN extinction correction, $f=0.86$);
\item f$_{\rm no-f}$ : Full sample with no correction factor applied
  (\Ko\,$ \le 11\fm75$, GN extinction correction, $f=1.0$);
\item f$_{\rm ext}$: Extended sample, with diameter-dependent extinction
  correction applied (\Kd\,$ \le 11\fm75$, GN extinction correction,
  $f=0.86$);
\item f$_{\rm ext-orig}$: Extended sample as in Paper I, with
  diameter-dependent extinction correction (\Kd\,$ \le 11\fm75$, SFD
  extinction correction, $f=0.83$).
\end{enumerate}

\item{Col.\ 8:} Galaxy disc type: `D1' means an obviously visible disc
  and/or spiral arms, `D2' stands for a noticeable disc, `D3' for a
  possible disc, and `D4' stands for no disc noticeable. Where it was not
  possible to tell (likely due to adjacent or superimposed stars or very
  high extinction) we give a flag `n'.

\item{Col.\ 9:} 2MRS flags: `c' stands for an entry in the main catalogue,
  and `e' for the extra catalogue (that is, outside the 2MRS selection
  criteria) . For the sub-catalogues\footnote{for a detailed description
    see http://tdc-www.harvard.edu/2mrs/2mrs\_readme.html} we give: `cf'
  for an entry in the flg catalogue (affected by nearby stars, but not
  severely), `cp' for an entry in the flr catalogue (severely affected by
  nearby stars), `ca' and `ea' are objects that have been reprocessed and
  are both in the rep as well as add catalogues (in the latter case with
  new IDs and photometry, not listed by us), `cr' and `er' are objects
  rejected as galaxies and which can be found in the rej catalogues, `cn'
  and `en' are objects that have no redshifts yet and are listed in the
  nocz catalogues.

\item{Cols. 10a -- 10c:} Velocity flags `vo', `vh' and `vtot': `o' stands
  for an optical measurement and `h' for \HI\ velocity; a colon indicates
  an uncertain measurement and a question mark stands for a questionable
  measurement, object ID or target (in particular for \HI : a radio
  telescope's large beam size may include nearby galaxies), or an uncertain
  detection (often with a low signal-to-noise ratio); `g' indicates a
  velocity measurement of a Galactic source\footnote{Though only objects
    flagged as galaxies (\cf Col.~7a) have velocity entries listed in the
    table, there are a few uncertain galaxies with questionable velocity
    measurements.}. Square brackets in Cols.~10a and 10b indicate a
  non-accepted velocity (due to questionable ID or measurement), and normal
  brackets in Col.~10c indicate not yet published measurements. In a few
  cases, though published values exist we prefer unpublished measurements
  (\eg when they have better errors).

\item{Col.\ 11:} Deep NIR image flag: `U' indicates a UKIDSS image for this
  object, `H' stands for a UHS, and `V' stands for VISTA.

\item{Col.\ 12:} Photometry flag: `O' indicates off-centre coordinates (\cf
  Col.~5), `A', `P' and `F' refer to the respective 2MRS flag (\cf Col.~9),
  and `C' denotes improved photometry exists in John Huchra's original list
  (see Paper I for details).

\item{Col.\ 13:} Extinction flag (see Paper I): `e' stands for a likely
  wrong extinction value, either due to small-scale variations or a
  non-removed FIR point source, `e?' denotes a possible wrong extinction
  value. Note that these have been updated since Paper I.

\item{Col.\ 14:} 2MASX magnitude $K_{\mathrm 20}$: isophotal \K -band
  magnitude measured within the \K -band 20 mag arcsec$^{-2}$ isophotal
  elliptical aperture (in magnitudes).

\item{Cols. 15 and 16:} 2MASX colours \hk\ and \jk : isophotal colours
  measured within the \K -band 20 mag arcsec$^{-2}$ isophotal elliptical
  aperture (in magnitudes).

\item{Col.\ 17:} 2MASX flag $vc$: the visual verification score of a
  source.

\item{Col.18:} 2MASX object size $a$: the major diameter of the object,
  which is twice the 2MASX \K -band 20 mag arcsec$^{-2}$ isophotal
  elliptical aperture semi-major axis {\tt r\_k20fe} (in arcseconds).

\item{Col.\ 19:} 2MASX axis ratio $b/a$: minor-to-major axis ratio fit to
  the $3\sigma$ super-co-added isophote, {\tt sup\_ba}. Single-precision
  entries are given in cases where the 2MASX parameter {\tt sup\_ba} is not
  determined; these values were estimated from the axial ratios available
  in the different passbands.

\item{Col.\ 20:} 2MASX stellar density: co-added logarithm of the number of
  stars (\K$<14$\,mag) per square degree around the object.

\item{Cols. 21 -- 23:} Magnitude \Ko\ and colours \hko\ and \jko\ corrected
  for foreground extinction (Col.~3) using a correction factor 0.86 (in
  magnitudes).

\item{Col.\ 24:} Extinction in the \K -band $\ak $: calculated from
  \ebv\ in Col.~3 and applying a correction factor of 0.86 (in magnitudes).

\item{Col.\ 25:} Extinction corrected major diameter $a^{\rm d}$ according
  to \citet{riad10} using \ebv\ from Col.~4 with the correction factor
  0.86; only given for objects classified as galaxies (in arcseconds).

\item{Col.\ 26:} Extinction corrected magnitude \Kd\ corrected according to
  \citet{riad10} using \ebv\ from Col.~3 with the correction factor 0.86;
  only given for objects classified as galaxies (in magnitudes).

\item{Cols. 27 and 28:} Adopted radial, heliocentric velocity with error
  and reference according to the vtot flag in Col.~10c. If the measurement
  is not yet published, we quote a token reference and do not give the
  velocity here.

\end{description}

\section{Properties of the catalogue } \label{prop}  

We have thoroughly discussed the properties of the bright 2MZoA catalogue
in Paper I, and most of it also applies to the full sample. In the
following, we show some updated plots and statistics. A sky map with the
distribution of the samples are shown in the appendix; it is not
qualitatively different from the maps shown in Paper I.

\begin{figure} 
\centering
\includegraphics[width=0.47\textwidth]{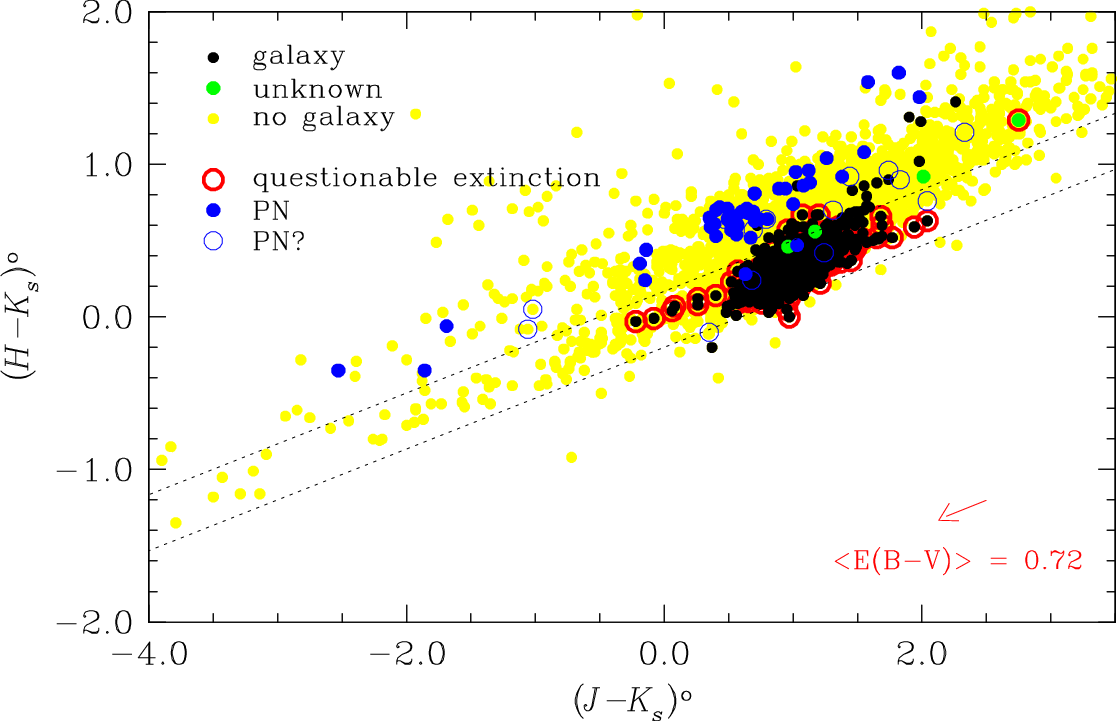}
\caption{2MASS extinction-corrected colour--colour plot, \hko\ versus \jko
  . A reddening vector, representing the mean of the sample shown, is
  indicated by a red arrow in the bottom right corner. The reddening path
  for objects within the intrinsic colour range of galaxies is indicated
  with two parallel dashed lines.
}
\label{colcolplot}
\end{figure}

Figure~\ref{colcolplot} shows the colour--colour plot with objects
classified as galaxies in black and non-galaxies in yellow; the few galaxy
candidates (objects of unknown classification, class 5) are shown in
green. Clearly offset in \hko\ colour are PNe, which are shown as blue dots
(where open circles indicate possible PNe). The reddening path for the
intrinsic colour range occupied by the bulk of the galaxies is shown by the
two dashed lines. Galaxies with wrong Galactic extinction values scatter
along this path. This is confirmed by the distribution of the red circles
which mark galaxies where we stated a questionable extinction based on
visually obvious variations across the images (extinction flag $e$).
Galaxies lying outside the reddening path are likely to be affected by
starlight in the photometry aperture.

\begin{figure} 
\centering
\includegraphics[width=0.47\textwidth]{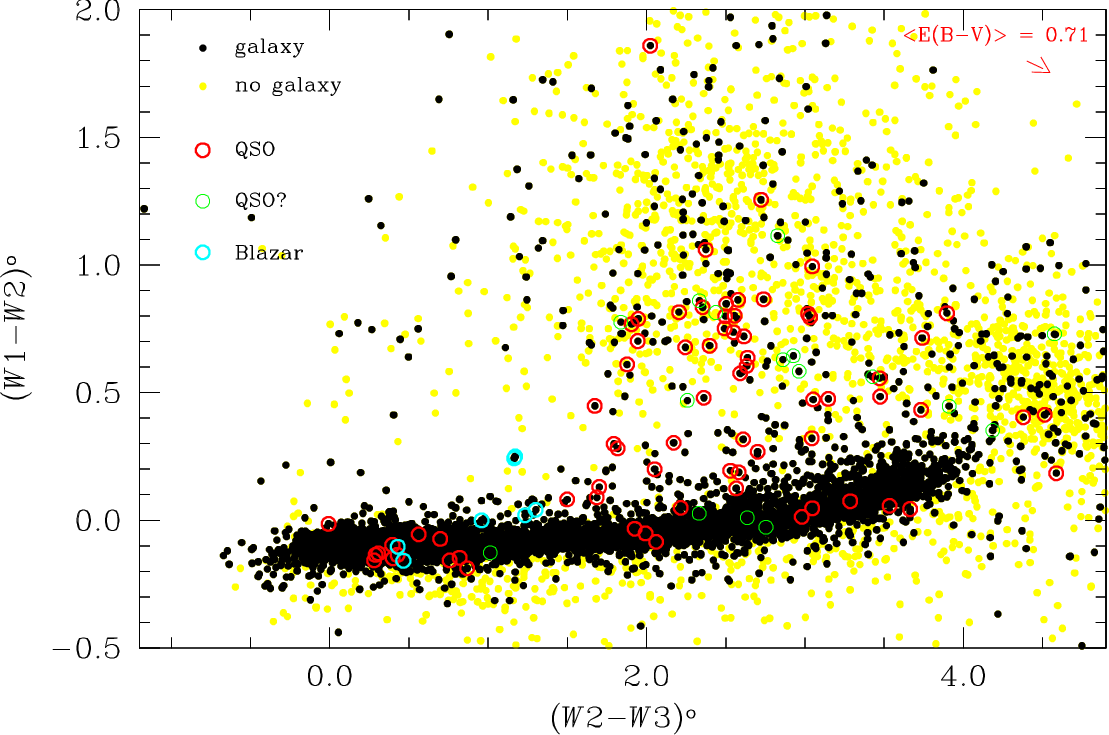}
\caption{WISE extinction-corrected colour--colour plot, $(W1-W2)^o$ versus
  $(W2-W3)^o$, for the full sample. A reddening vector, representing the
  mean of the sample shown, is indicated by a red arrow in the top right
  corner. Open circles mark active galaxies as labelled.
}
\label{wisecolplot}
\end{figure}

For the WISE colour--colour plot (Fig.~\ref{wisecolplot}) we use the
aperture 1 ($5\farcs5$) magnitudes of the ALLWISE catalogue, corrected for
foreground extinction, as in Paper I. Spiral galaxies are located more to
the right, and ellipticals are predominantly found to the left (see
Figure~26 in \citealt{jarrett11}), while active and intensely star forming
galaxies are separated in ($W1-W2$), top. We marked all AGNs listed in the
literature with red circles and blazars with cyan circles; possible AGNs
(based on their colour and visual appearance) are marked with green
circles.

\begin{figure} 
\centering
\includegraphics[width=0.35\textwidth]{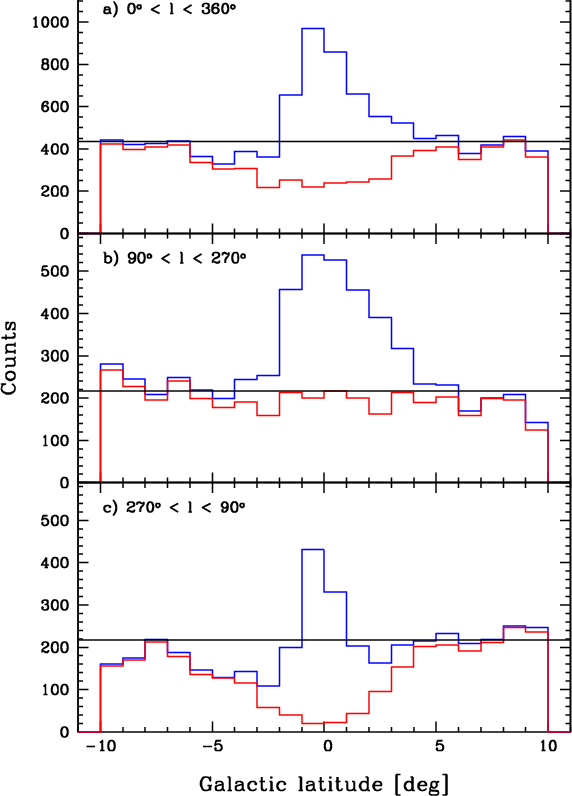}
\caption{Histograms of objects in the ZOA sample as a function of Galactic
  latitude for all objects (blue) and for galaxies only (red). The black
  line represents the average 2MASX galaxy count outside the ZoA. Panel
  (a): for all Galactic longitudes; panel (b) for the semi-circle towards
  the anti-centre region; panel (c) for the semi-circle towards the bulge.
}
\label{galbhistplot}
\end{figure}

The completeness of a sample of galaxies can be shown in various
ways. Figure~\ref{galbhistplot} shows histograms of source counts as a
function of Galactic latitude and longitude. While the galaxy counts (red)
show a small dip near the Galactic plane (Panel (a)), the distribution of
all sources (in blue) shows a clear peak, emphasising the need for visual
inspection to identify non-galaxian extended objects in the 2MASX catalogue
for the inner ZoA ($|b|<5\degr$). For comparison, the black line represents
the average galaxy density outside of the ZoA, which is 1.212
galaxies/sq.deg (black line) based on source counts of the 2MASX catalogue
with $|b|>30\degr$ and $K_{\rm 20} \le 11\fm75$. Panels (b) and (c) show
the semi-circles towards the anti-centre and the Galactic bulge,
respectively. The dip in galaxy counts disappears in the anti-centre
direction but is greatly pronounced towards the Galactic bulge. As already
noted in Paper I, the dip towards the Galactic Bulge indicates a breakdown
in galaxy recognition of the 2MASX survey mainly due to star crowding.

\begin{figure} 
\centering \includegraphics[width=0.35\textwidth]{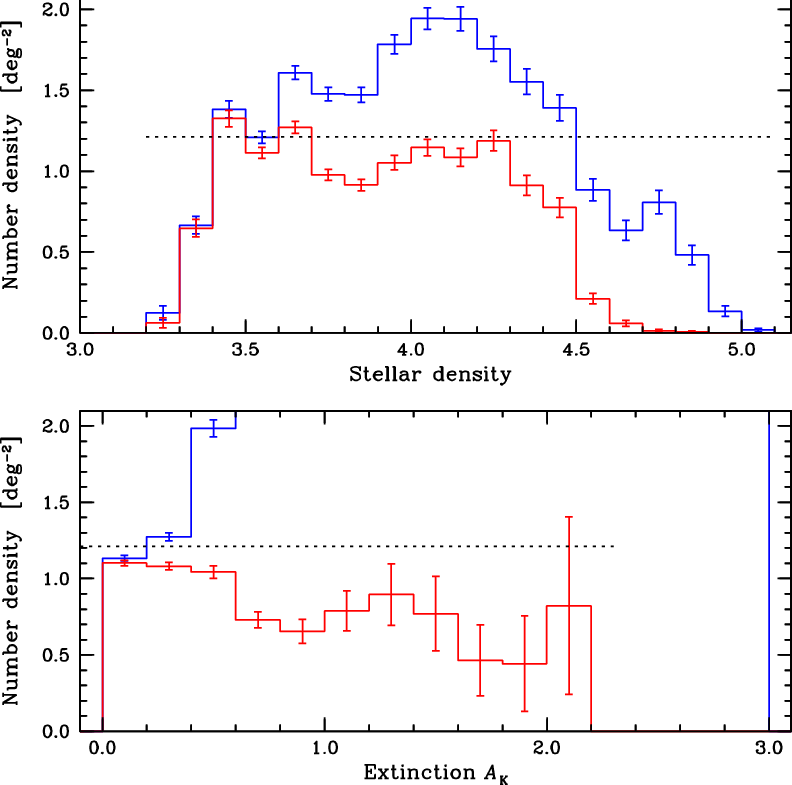}
\caption{Histograms of the number density of all ZOA sample objects (blue)
  and galaxies only (red) as a function of stellar density (top panel) and
  Galactic extinction $\ak $ (bottom panel; only regions with $\log
  N_*/{\rm deg}^2 \le 4.5$ were used). The error bars indicate Poissonian
  errors per bin. The black dotted lines represent the average 2MASX bright
  galaxy number density outside of the ZoA.
}
\label{histextstdplot}
\end{figure}

We can also show completeness levels as a function of stellar densities and
extinction, see Fig.~\ref{histextstdplot} top and bottom panel,
respectively.  For the extinction plot we only used areas with stellar
densities $\log N_*/{\rm deg}^2 <4.5$ (to avoid a selection bias since both
high extinction and high stellar density occur in similar areas).  The
number densities are highly affected by low number statistics for $\ak \,
\approxgt \, 1\fm0$. Due to the small area covered by our sample, the
numbers are also subject to cosmic variance. However, we can make two broad
statements: Extinction seems to affect galaxy counts at all levels, and the
effect of stellar density is negligible up to $\log N_*/{\rm deg}^2 \sim
4.3$ and becomes severe at $\log N_*/{\rm deg}^2 > 4.5$.

Table~\ref{tabparam} gives the statistics on some of the parameters in the
ZOA and EBV catalogues. For better comparison, we give percentages with
respect to all galaxies in the respective sample, with the total numbers
listed at the bottom. Regarding the disc parameter, we find that over 80\%
of all galaxies show a disc, while only around 4\% have no indication of
one.  Extinction flags were set for $\sim\!20$\% of galaxies in the ZOA
sample, with a much higher fraction of $\sim\!65$\% for the EBV sample. The
latter is expected since the EBV sample covers only high extinction areas.

\begin{table}  
\centering
\begin{minipage}{140mm}
\caption{{Statistics on some parameters for both samples. } \label{tabparam}}
{\small
\begin{tabular}{llrr}
\hline
\noalign{\smallskip}
Param & meaning     & ZOA sample & EBV sample \\
      &             & [\%] & [\%] \\
\noalign{\smallskip}
\hline
\noalign{\smallskip}
\multicolumn{4}{l}{Disc class: }  \\
\noalign{\smallskip}
D1  & obvious disc   & 55.1 &  54.2 \\  
D2  & disc           & 25.4 &  29.6 \\  
D3  & possible disc  & 13.6 &   9.2 \\  
D4  & no disc        &  4.3 &   3.5 \\  
n   & unable to tell &  0.7 &   0.7 \\  
\noalign{\smallskip}
\hline
\noalign{\smallskip}
\multicolumn{4}{l}{Extinction flag: }  \\
\noalign{\smallskip}
$e$           & likely wrong ext   & 11.7  & 44.4 \\  
$e$?          & possibly wrong ext &  9.6  & 19.0 \\  
\hline
\noalign{\smallskip}
\multicolumn{4}{l}{Velocities: }  \\
\noalign{\smallskip}
$v_{\rm tot}$ & adopted (opt or HI) & 72.4  & 76.7 \\  
$v_{\rm op}$  & opt                 & 64.0  & 70.4 \\  
$v_{\rm op}$? & uncertain ID        &  0.3  &  1.4 \\  
$v_{\rm hi}$  & \HI\                & 17.0  & 21.8 \\  
$v_{\rm hi}$? & uncertain ID        &  0.3  &  0.7 \\  
$v_{\rm lit-op}$  & published       & 58.0  & 68.3 \\  
$v_{\rm lit-op}$? & uncertain ID    &  0.5  &  7.0 \\  
$v_{\rm lit-hi}$  & published       & 16.1  & 17.6 \\  
$v_{\rm lit-hi}$? & uncertain ID    &  1.2  &  0.7 \\  
2MRS          & main catalogue      & 49.8  & 16.2 \\  
2MRS          & extra catalogue   & 10.8  & 54.2 \\  
\noalign{\smallskip}
\hline\hline
\noalign{\smallskip}
\multicolumn{2}{l}{Total $N$} & 6757 & 142 \\
\noalign{\smallskip}
\hline
\end{tabular}
}
\end{minipage}
\end{table}

As to redshift information, more than 70\% of the sample galaxies have
redshift measurements, the majority of which are optical. We distinguish
between optical and \HI\ measurements, and also whether the velocity
information has been published\footnote{Some of the redshift measurements
  we use in our catalogue are work in progress and were obtained by members
  of our group and collaborators, see \citet{schroeder21} for details.}
($v_{\rm lit}$). We separately list cases where it appears questionable
that the measured redshift belongs to the targeted galaxy (indicated with a
question mark). This mainly concerns \HI\ measurements where the often very
large beam size can include other galaxies close by. Though we give such
velocities in our catalogue, they have been flagged in Col.~10c with a
question mark and are ignored in our analysis.

We have also flagged all objects that appear in the 2MRS catalogues
(\citealt{huchra12}, \citealt{macri19}), though the percentages given here
relate only to objects classified by us as galaxies.  Furthermore, we have
noted but not extracted photometric redshift information in the
literature\footnote{The majority of which can be found in the 2MASS
  Photometric Redshift catalogue by \citet{bilicki14}.}. Since photometric
redshifts have large uncertainties (and can be affected by stars in the
aperture), we do not find them useful for our, predominantly local,
sample. Finally, we noted a number of possible \HI\ detections, which need
confirmation before we can make use of them.

\begin{table*}
\centering
\begin{minipage}{280mm}
\caption{{Statistics on 2MASX colours of various samples} \label{tab2mhistb}}
{\scriptsize
\begin{tabular}{l@{\extracolsep{2.0mm}}c@{\extracolsep{2.0mm}}c@{\extracolsep{2.0mm}}c@{\extracolsep{2.0mm}}c@{\extracolsep{0.0mm}}c@{\extracolsep{2.0mm}}c@{\extracolsep{2.0mm}}c@{\extracolsep{2.0mm}}c@{\extracolsep{2.0mm}}c@{\extracolsep{0.0mm}}c}
\noalign{\smallskip}
\hline\hline
\noalign{\smallskip}
Sample       & \multicolumn{5}{c}{\jk } & \multicolumn{5}{c}{\hk } \\
\noalign{\smallskip}
\cline{2-6} \cline{7-11} 
\noalign{\smallskip}
       & $N$   & median & mean & error & standard deviation & $N  $ &  median & mean & error & standard deviation \\
       &       & [mag]  & [mag]& [mag] & [mag]              &       &    [mag]& [mag]& [mag] & [mag]              \\
\noalign{\smallskip}
\hline
\noalign{\smallskip}
Colour$^{\rm o}$\,     &    4932 & $0.990$ & $1.001$ & $0.002$ & $0.11$ &    5000 & $0.300$ & $0.306$ & $0.001$ & $0.08$ \\
Colour$^{\rm o,c}$\,   &    4904 & $1.011$ & $1.019$ & $0.002$ & $0.11$ &    4972 & $0.309$ & $0.318$ & $0.001$ & $0.08$ \\
Colour   (high-lat)    & 24\,966 & $1.018$ & $1.030$ & $0.001$ & $0.13$ & 24\,967 & $0.306$ & $0.315$ & $0.001$ & $0.09$ \\
Colour   (high-lat)    &$2952-3290$ & $1.009-1.027$ & $1.019-1.044$ & $0.001-0.002$ & $0.10-0.20$ & $2952-3289$ & $0.301-0.311$ & $0.308-0.323$ & $0.001-0.002$ & $0.07-0.13$ \\
\noalign{\smallskip}
\hline
\end{tabular}
}
\end{minipage}
\end{table*}

\begin{figure} 
\centering
\includegraphics[width=0.4\textwidth]{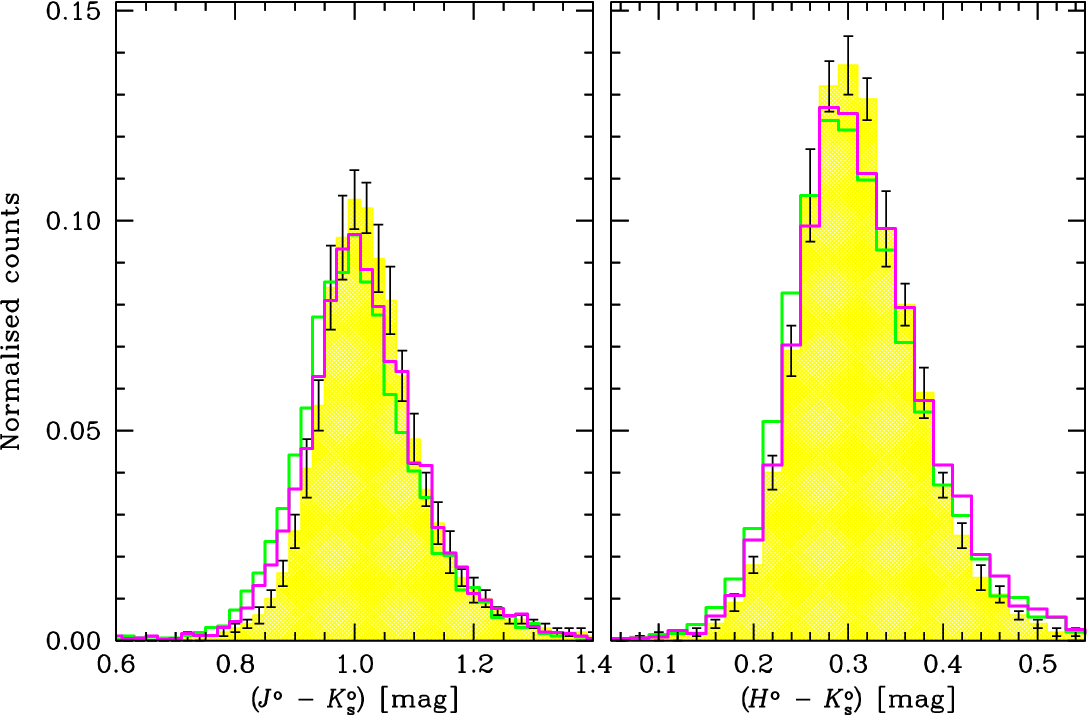}
\caption{Normalised histograms of 2MASX colours \jko\ (left) and
  \hko\ (right). Green histograms are the full sample and the magenta
  histograms show colours corrected for stellar densities. the
  yellow-filled histograms represent the average high Galactic latitude
  sample.
}
\label{2mhist2}
\end{figure}

Figure~\ref{2mhist2} shows the histograms for the \jko\ and \hko\ colours
(green lines). In Paper I we had found that there is a dependence on
stellar density. We re-analysed the relationship for the new sample and
found slightly improved dependencies:
\[(J-K_{\rm s})^{\rm o} = (-0.064\pm0.007) \cdot \log N_*/{\rm deg}^2 + (1.244\pm0.027) \] 
and
\[(H-K_{\rm s})^{\rm o} = (-0.039\pm0.005) \cdot \log N_*/{\rm deg}^2 + (0.453\pm0.020) .\] 
The magenta-coloured histogram shows the accordingly corrected colours, and
Table~\ref{tab2mhistb} gives the statistics of the colours with and without
the correction. As in Paper I, we also show the histograms for galaxies
outside the ZoA for comparison (yellow). Contrary to Paper I, we choose a
higher latitude cut-off to make sure that extinction will not affect the
(uncorrected) colours. With cuts at $|b| > 30\degr$ and \K $<11\fm75$, the
high-latitude sample comprises 25\,007 galaxies. As described in Paper I in
more detail, we formed 16 subsamples containing each about 3000 galaxies to
test for cosmic variance. The yellow histograms show the mean normalised
counts with error bars based on the standard deviation of the 16
subsamples. The mean as well as the range of values for the 16
high-latitude subsamples are also given in Table~\ref{tab2mhistb}. Since
colours are affected by $k$-correction (noticeable as an extended red tail
in the histograms), we prefer the median to the mean. There is good
agreement between the high and low latitude samples if we apply the stellar
density correction to our sample.

\section{Sample comparisons }\label{comp}  

There are two major differences to Paper I:
\begin{enumerate}
\item As explained above, we use an improved extinction correction, which
  changes the composition of the sample for \Ko\,$ \le 11\fm25$. In
  addition, with new deep NIR images and velocity information, some objects
  have a changed object class and have thus entered or left the bright
  galaxy sample.

\item By extending the magnitude limit from $11\fm25$ to $11\fm75$, the
  qualitative composition of the sample likely has changed since fainter
  galaxies are either smaller or more distant galaxies.
\end{enumerate}

\subsection{Comparison with Paper I}

We can make a direct comparison with Paper I by applying the same magnitude
limit $11\fm25$ to our here presented sample. First of all, the new sample
is smaller since we also apply a correction factor smaller than one to the
extinction values, resulting in fainter magnitudes, which affects the
sample size. For clarity, we have also compared the effects of applying the
SFD corrections (with and without a correction factor) instead of the GN
extinction. The top section of Table~\ref{tabsamples2} gives the various
sample sizes.

\begin{table}
\caption[]{Various samples with different \Ko\ band limits.
\label{tabsamples2}}
{\tiny
\begin{tabular}{l@{\extracolsep{-2.0mm}}clrrrr}
\hline \hline
\noalign{\smallskip}
Sample    &   Source of & $f$ & \multicolumn{2}{c}{ZOA sample} & \multicolumn{2}{c}{EBV sample} \\
          & extinction  &  & $N_{\rm obj}$ & $N_{\rm gal}$ & $N_{\rm obj}$ & $N_{\rm gal}$ \\
\noalign{\smallskip}
\hline
\noalign{\smallskip}
\multicolumn{7}{l}{Bright samples ($K^{\rm o}_{\rm s} \le 11\fm25$):} \\
Paper I                   & SFD & 1.0  & 6913 & 3693 & 502 & 89 \\  
This paper (S1)           & GN  & 0.86 & 6080 & 3470 & 424 & 76 \\  
$-$ Variation A           & SFD & 0.83 & 6448 & 3433 & 465 & 78 \\  
$-$ Variation B           & SFD & 1.0  & 6835 & 3645 & 492 & 88 \\  
$-$ Variation C           & GN  & 1.0  & 6393 & 3655 & 441 & 78 \\  
\noalign{\smallskip}
\hline 
\noalign{\smallskip}
\multicolumn{7}{l}{Fainter samples:} \\
S2 ($11\fm25 < K^{\rm o}_{\rm s} \le 11\fm50$) & GN & 0.86 & 1577 & 1305 & 53 & 20 \\ 
S3 ($11\fm50 < K^{\rm o}_{\rm s} \le 11\fm75$) & GN & 0.86 & 2282 & 1982 & 77 & 46 \\ 
\noalign{\smallskip}
\hline
\noalign{\smallskip}
\multicolumn{7}{l}{Supplementary sample:} \\
S$+$ (\Ko\,$ > 11\fm75$,  \Kd\,$ \le 11\fm75$) & GN & 0.86 & 1021 &  587 & 62 & 18 \\
\noalign{\smallskip}
\hline
\end{tabular}
}
\end{table}

While for each galaxy the GN extinction value may be smaller or larger than
the SFD value, the overall effect on the sample is that we lose more
objects than we gain. However, this seems to be only the case for the
non-galaxies ($\Delta N = 405$ and $452$ for the ZOA sample, depending on
whether $f$ was applied or not, respectively), while the differences in the
galaxy samples are negligible ($\Delta N = -37$ and $-10$,
respectively). We therefore do not expect (and have not found) any
qualitative differences with the discussion in Paper I regarding the bright
galaxy sample.

\subsection{Bright versus faint } \label{faint}

To investigate qualitative differences between bright and faint galaxies,
we have divided our sample into the three subsamples bright (S1), medium
(S2) and faint (S3), with \Ko\,$ \le 11\fm25$, $11\fm25 < K^{\rm o}_{\rm s}
\le 11\fm50$ and $11\fm50 < K^{\rm o}_{\rm s} \le 11\fm75$, respectively.
Sample sizes are given in the second part of Table~\ref{tabsamples2}.

\begin{figure} 
\centering \includegraphics[width=0.48\textwidth]{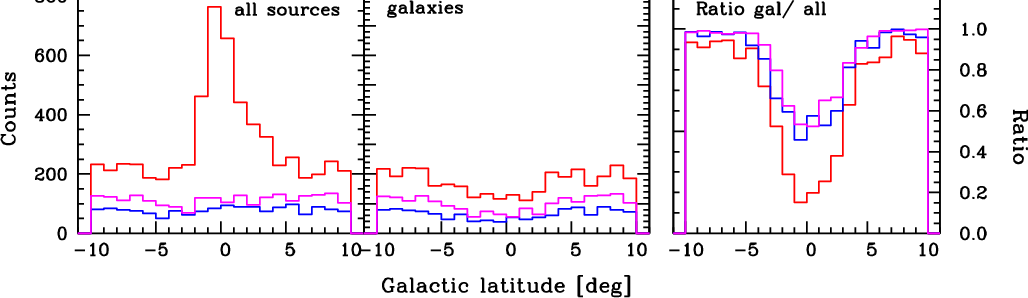}
\caption{Histograms of objects in the ZOA samples with Galactic latitude:
  all objects (left), galaxies (middle) and the fraction of galaxies
  (right). S1 is shown in red, S2 in blue and S3 in magenta).  }
\label{galbhist2plot}
\end{figure}

\begin{figure} 
\centering \includegraphics[width=0.48\textwidth]{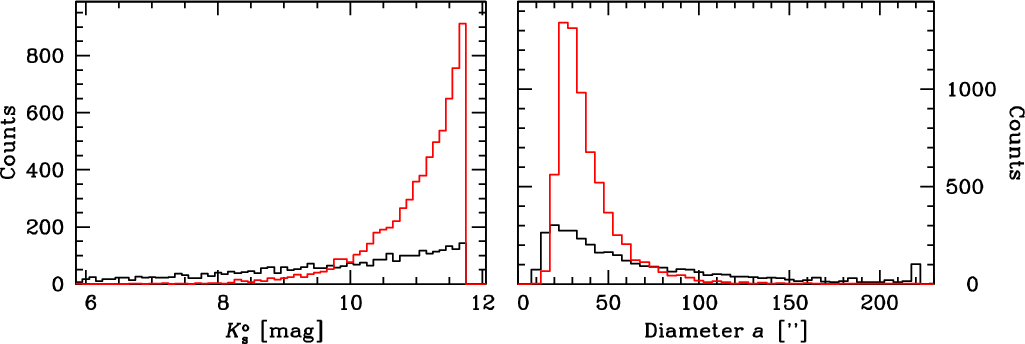}
\caption{Histograms of objects in the full sample: non-galaxies are in
  black and galaxies in red. Left: As a function of extinction-corrected
  \Ko\ band magnitude; right: as a function of diameter in arcseconds.
}
\label{histko2plot}
\end{figure}

First of all, we show the distribution of the three subsamples across
Galactic latitudes in Fig.~\ref{galbhist2plot} for all sources (left
panel), galaxies (middle panel) and the ratio of galaxies versus all
sources (right panel). While the dip in galaxy numbers seen in
Fig.~\ref{galbhistplot} around the Galactic plane persists (and varies
similarly with longitude, not shown), only the S1 sample (red) shows a
clear peak of non-galaxies near the plane. To investigate this further, we
show the \Ko\ band histogram in Fig.~\ref{histko2plot} (left panel) for
non-galaxies in black and galaxies in red. While the galaxies show the
typical exponential luminosity function, the number counts of non-galaxies
increases much slower (almost but not quite linearly). It should be noted
that the non-galaxies are mainly Galactic nebulae or parts thereof, with
some artefacts and stars or batches of faint stars added to the mix. Due to
their size and the overestimation of foreground extinction, the detections
of the nebulae are usually bright, and it is therefore obvious that the
fainter samples show much less contamination. To emphasise this, we show
the histogram of diameters (right hand panel in Fig.~\ref{histko2plot}). At
large diameters the non-galaxies dominate as expected. Non-galaxies also
dominate at the smallest diameters, which is likely due to the spurious
detection of stars.

\begin{figure} 
\centering \includegraphics[width=0.48\textwidth]{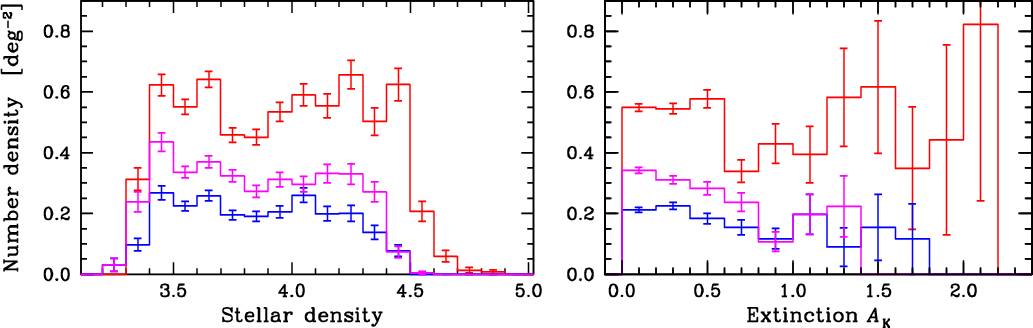}
\caption{Same as Fig.~\ref{histextstdplot} for the galaxies in the
  subsamples S1 (red), S2 (blue) and S3 (magenta).
}
\label{histextstd2plot}
\end{figure}

Figure~\ref{histextstd2plot} compares the samples regarding stellar density
(left panel) and extinction (right panel). All three samples show stable
number densities for stellar densities of $\log N_*/{\rm deg}^2 \ge 3.4$,
the sharp cut-off at $\log N_*/{\rm deg}^2 \ge 4.5$ known from Paper I (and
here visible for S1) is less pronounced for the fainter samples with a
reduced limit of $\log N_*/{\rm deg}^2 \sim 4.4$, indicating that stellar
densities affect more the detection of smaller and fainter galaxies. The
extinction also affects more strongly the two fainter samples.  Due to the
limit in extinction-corrected magnitude, the maximum $\ak $ value drops
from $2\fm1$ to $1\fm7$ to $1\fm3$, respectively. The faintest sample S3
(magenta) shows a decrease even in the first three bins ($\ak < 0\fm6$) and
both S2 and S3 show a gradual decrease beyond $\ak = 0\fm6$.  We conclude
that the faintest sample is not complete at any extinction level.

Statistics of the parameters disc class, extinction flag and availability
of redshift measurement information are presented in Table~\ref{tabparam2},
similarly to Table~\ref{tabparam}. While the disc class parameter shows the
increasing difficulties to classify fainter (and smaller) galaxies, the
determination of the extinction flag is independent of the kind of detected
galaxy. There is also a marked (and expected) decrease in redshift
measurements for the fainter samples, especially in \HI\ measurements. This
is partly because mostly bright and large galaxies are targeted in
\HI\ surveys. In addition, the sensitivity decreases rapidly with distance
so that most blind \HI\ surveys do not go much beyond $v=10\,000$\,\kms
. However, this will change with the new generation of radio telescopes
like MeerKAT, ASKAP and, eventually, the SKA.

Representation in the 2MRS catalogue is stable across all samples as
expected since we choose the same magnitude limit. The inclusion in the
2MRS extra catalogue decreases for the fainter samples since galaxies were
not systematically included in this catalogue but only through availability
of redshift measurements in the literature.

\begin{table}  
\centering
\begin{minipage}{140mm}
\caption{{Statistics on some parameters for the subsamples. } \label{tabparam2}}
{\tiny
\begin{tabular}{lrrrr@{\extracolsep{6.0mm}}r@{\extracolsep{2.0mm}}r@{\extracolsep{2.0mm}}r@{\extracolsep{2.0mm}}r}
\hline
\noalign{\smallskip}
Param & \multicolumn{4}{c}{ZOA samples} & \multicolumn{4}{c}{EBV samples}  \\
      & S1 & S2 & S3 & S+ & S1 & S2 & S3 & S+ \\  
      & [\%] & [\%] & [\%] & [\%]& [\%] & [\%] & [\%] & [\%] \\
\noalign{\smallskip}
\hline
\noalign{\smallskip}
\multicolumn{9}{l}{Disc class: }  \\
\noalign{\smallskip}
D1  & 59.0 & 54.8 & 48.4 & 43.3 & 55 & 50  & 54 & 61 \\ 
D2  & 21.8 & 26.0 & 31.3 & 28.4 & 33 & 35  & 22 & 11 \\ 
D3  & 11.0 & 15.1 & 17.1 & 19.8 &  5 & 10  & 15 & 17 \\ 
D4  &  6.1 &  2.9 &  2.2 &  3.7 &  3 &  5  &  4 & 11 \\ 
n   &  0.3 &  1.1 &  1.1 &  4.8 &  1 &  0  &  0 &  0 \\ 
\noalign{\smallskip}
\hline
\noalign{\smallskip}
\multicolumn{9}{l}{Extinction flag: }  \\
\noalign{\smallskip}
$e$           & 12.2 & 11.2 & 11.3 & 20.4 & 38 & 45 & 54 & 44 \\ 
$e$?          &  9.2 &  9.0 & 10.2 & 19.8 & 17 & 20 & 22 & 28 \\ 
\hline
\noalign{\smallskip}
\multicolumn{9}{l}{Velocities: }  \\
\noalign{\smallskip}
$v_{\rm tot}$     & 85.5 & 65.0 & 54.7 & 14.8 &  93 & 75 & 50 & 22  \\ 
$v_{\rm op}$      & 72.9 & 59.0 & 51.8 & 13.3 &  86 & 75 & 44 & 17  \\ 
$v_{\rm op}$?     &  0.4 &  0.3 &  0.3 &  0   &   0 &  5 &  2 &  0  \\ 
$v_{\rm hi}$      & 26.6 & 10.7 &  4.6 &  1.9 &  30 & 20 &  9 &  6  \\ 
$v_{\rm hi}$?     &  0.3 &  0.5 &  0.3 &  0.2 &   0 &  0 &  2 &  0  \\ 
$v_{\rm lit-op}$  & 62.7 & 55.1 & 51.7 & 13.3 &  83 & 70 & 44 & 17  \\ 
$v_{\rm lit-op}$? &  0.7 &  0.5 &  0.3 &  0.2 &   4 & 10 & 11 &  6  \\ 
$v_{\rm lit-hi}$  & 25.0 & 10.3 &  4.6 &  1.9 &  26 & 15 &  4 &  6  \\ 
$v_{\rm lit-hi}$? &  1.4 &  0.9 &  0.7 &  0.2 &   0 &  0 &  2 &  0  \\ 
2MRS main         & 49.3 & 50.2 & 50.6 & 11.4 &  20 &  5 & 15 & 17  \\ 
2MRS extra        & 17.4 &  6.1 &  2.6 &  1.7 &  66 & 65 & 30 &  0  \\ 
\noalign{\smallskip}
\hline\hline
\noalign{\smallskip}
\multicolumn{1}{l}{Total $N$} & 3470 & 1305 & 1982 & 587 & 76 & 20 & 46 & 18 \\ 
\noalign{\smallskip}
\hline
\end{tabular}
}
\end{minipage}
\end{table}

\begin{table*}
\centering
\begin{minipage}{280mm}
\caption{{Statistics on 2MASX colours of various samples. 
} \label{tab2mhistb2}}
{\scriptsize
\begin{tabular}{l@{\extracolsep{2.0mm}}c@{\extracolsep{2.0mm}}c@{\extracolsep{2.0mm}}c@{\extracolsep{2.0mm}}c@{\extracolsep{0.0mm}}c@{\extracolsep{2.0mm}}c@{\extracolsep{2.0mm}}c@{\extracolsep{2.0mm}}c@{\extracolsep{2.0mm}}c@{\extracolsep{0.0mm}}c}
\noalign{\smallskip}
\hline\hline
\noalign{\smallskip}
Sample       & \multicolumn{5}{c}{\jk } & \multicolumn{5}{c}{\hk } \\
\noalign{\smallskip}
\cline{2-6} \cline{7-11} 
\noalign{\smallskip}
       & $N$   & median & mean & error & standard deviation & $N  $ &  median & mean & error & standard deviation \\
       &       & [mag]  & [mag]& [mag] & [mag]              &       &    [mag]& [mag]& [mag] & [mag]              \\
\noalign{\smallskip}
\hline
\noalign{\smallskip}
Colour$^{\rm o}$\, S1   &    2547 & $0.980$ & $0.985$ & $0.002$ & $0.10$ &    2582 & $0.290$ & $0.292$ & $0.001$ & $0.07$ \\
Colour$^{\rm o,c}$\, S1 &    2520 & $0.994$ & $1.004$ & $0.002$ & $0.10$ &    2555 & $0.298$ & $0.305$ & $0.001$ & $0.07$ \\
Colour   (high-lat) S1  & 12\,607 & $1.000$ & $1.013$ & $0.001$ & $0.13$ & 12\,609 & $0.293$ & $0.302$ & $0.001$ & $0.09$ \\
Colour   (high-lat) S1  &$1444-1719$ & $0.991-1.011$ & $1.002-1.028$ & $0.002-0.004$ & $0.10-0.23$ & $1441-1718$ & $0.289-0.298$ & $0.295-0.310$ & $0.001-0.003$ & $0.06-0.14$ \\
Colour$^{\rm o}$\, S2   &     965 & $1.010$ & $1.011$ & $0.004$ & $0.11$ &     976 & $0.310$ & $0.314$ & $0.002$ & $0.08$ \\
Colour$^{\rm o,c}$\, S2 &     965 & $1.025$ & $1.028$ & $0.004$ & $0.11$ &     976 & $0.320$ & $0.325$ & $0.002$ & $0.08$ \\
Colour   (high-lat) S2  &    5156 & $1.030$ & $1.042$ & $0.002$ & $0.12$ &    5156 & $0.315$ & $0.322$ & $0.001$ & $0.08$ \\
Colour   (high-lat) S2  &$601-697$& $1.021-1.041$ & $1.031-1.058$ & $0.003-0.007$ & $0.09-0.20$ & $601-697$ & $0.310-0.321$ & $0.315-0.332$ & $0.001-0.002$ & $0.06-0.12$ \\
Colour$^{\rm o}$\, S3   &    1420 & $1.020$ & $1.025$ & $0.003$ & $0.12$ &    1442 & $0.320$ & $0.325$ & $0.002$ & $0.10$ \\
Colour$^{\rm o,c}$\, S3 &    1419 & $1.034$ & $1.041$ & $0.003$ & $0.12$ &    1441 & $0.327$ & $0.336$ & $0.002$ & $0.09$ \\
Colour   (high-lat) S3  &    7170 & $1.042$ & $1.052$ & $0.001$ & $0.11$ &    7169 & $0.325$ & $0.332$ & $0.001$ & $0.08$ \\
Colour   (high-lat) S3  &$835-949$& $1.027-1.054$ & $1.042-1.066$ & $0.002-0.004$ & $0.09-0.14$ & $835-950$ & $0.317-0.333$ & $0.324-0.340$ & $0.002-0.003$ & $0.07-0.10$ \\
\noalign{\smallskip}
\hline
\end{tabular}
}
\end{minipage}
\end{table*}

Finally, we present the mean and median colours for the various samples in
Table~\ref{tab2mhistb2}. It is obvious that the colours become redder for
the fainter samples, which is also the case for the high-latitude samples
and thus not caused by the higher extinctions. Instead, such a shift is
expected since we have not applied a $k$-correction. A similar shift has
been noted for the bright sample in Paper II (see their Figure~8) where we
compared (uncorrected) colours of galaxies in various redshift bins.

\subsection{Extended sample}

We have also derived a sample by applying a diameter-dependent extinction
correction as explained above (and, in more detail, in Paper I). Since this
is an additional correction, our sample is not complete anymore and we have
to add those objects which have \Ko\,$ > 11\fm75$ but \Kd\,$ \le
11\fm75$. The supplementary sample, labelled S$+$, comprises 1083 objects
(see last entry in Table~\ref{tabsamples2}) and increases the galaxy sample
by about 9\%. The combined sample is called extended sample or S$e$.

\begin{figure} 
\centering \includegraphics[width=0.5\textwidth]{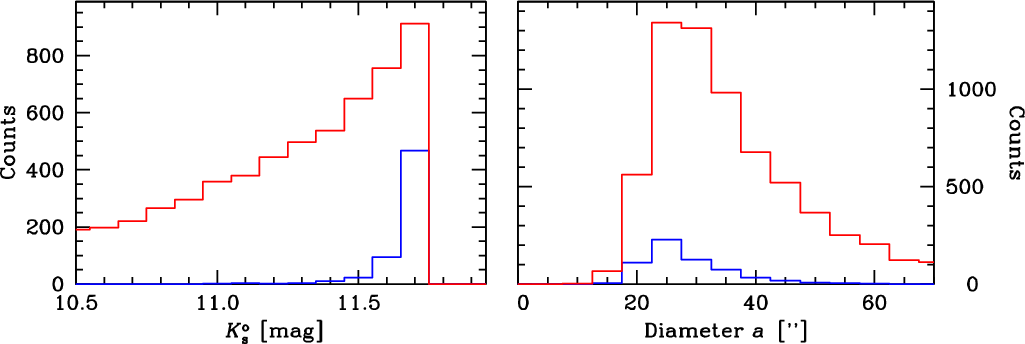}
\caption{Histograms of galaxies in the supplementary sample (S$+$, blue) as
  compared to the full sample (S, red). Left: As a function of
  extinction-corrected \K\ band magnitude (with the diameter-dependent
  extinction correction applied for the S$+$ sample); right: as a function
  of diameter in arcseconds (also corrected for the S$+$ sample). The red
  histograms are truncated.
}
\label{histko3plot}
\end{figure}

The galaxies in the supplementary sample are faint, with a lower limit of
\Kd\,$ =11\fm00$ as shown in Fig.~\ref{histko3plot} in blue. For
comparison, the full sample is shown in red. Note that we use the
appropriately corrected \K\ band magnitude for each sample. We also show
the distribution with diameter, with a maximum diameter $a^{\rm d} =
61\arcsec$ for S+.

\begin{figure} 
\centering \includegraphics[width=0.5\textwidth]{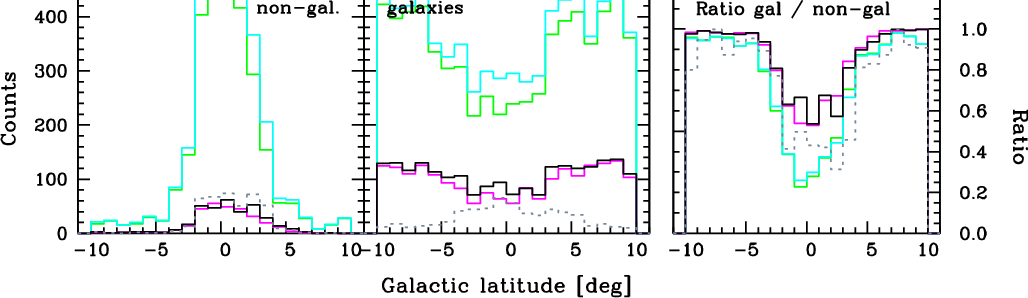}
\caption{Same as Fig.~\ref{galbhist2plot} for the full sample S (green),
  S$e$ (cyan), S3 (magenta), S3$e$ (the faint subsample of the extended
  sample; black) and S+ (dotted grey). The left-hand panel shows
  non-galaxies instead of all sources and is truncated to emphasise the
  comparison of the faint samples.
}
\label{galbhist3plot}
\end{figure}

\begin{figure} 
\centering \includegraphics[width=0.48\textwidth]{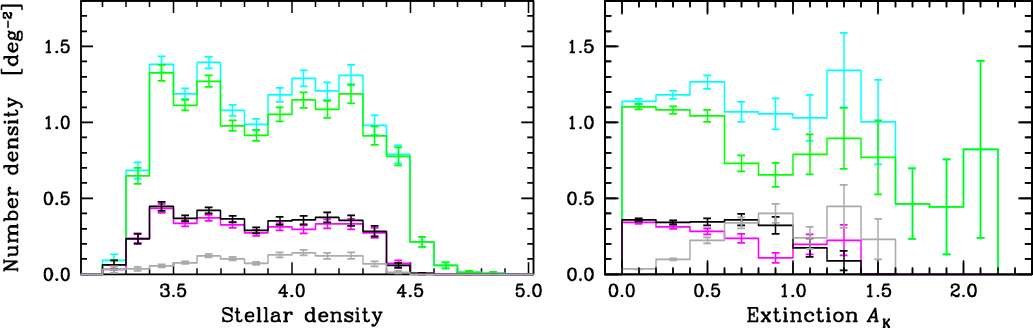}
\caption{Same as Fig.~\ref{histextstd2plot} for the full sample S (green),
  S$e$ (cyan), S3 (magenta), S3$e$ (the faint subsample of the extended
  sample; black) and S+ (dotted grey).
}
\label{histextstd3plot}
\end{figure}

The histograms over Galactic latitude, stellar densities and extinction are
shown in Figures~\ref{galbhist3plot} and~\ref{histextstd3plot}. Here we
compare, on the one hand, the full sample S with the extended sample S$e$
(in green and cyan, respectively), and, on the other hand, the two faint
subsamples S3 and S3$e$ with corrected magnitudes in the range $11\fm50 -
11\fm75$ (in magenta and black, respectively). For interest, we have added
the S+ sample in dotted grey line. The differences between the two faint
samples seems small, but the distribution of the S+ galaxies with a marked
peak around the Galactic plane shows the importance of this sample for the
completion rate. This is more pronounced in Fig.~\ref{histextstd3plot}
where the S+ distribution with extinction shows a clear increase, resulting
in the S3$e$ sample to show a more even distribution with extinction than
S3. We estimate the completion in the S$e$ sample is about $\ak = 1\fm3$
and possibly up to $\ak = 1\fm5$. The distribution with stellar densities
appears unaffected.

We added the statistics on the S+ sample to Table~\ref{tabparam2} though
the total numbers of galaxies (587 and 18 in the ZOA and EBV samples,
respectively) are small. Nonetheless, we can confirm that the trends with
fainter samples hold for Disc class and velocities (with only 87 and 4
galaxies having a velocity measurement, respectively), while the extinction
flags show a marked increase of almost a factor two for the ZOA sample.

Colours are largely unaffected by the additional diameter-dependent
extinction correction\footnote{Colours are derived using a fixed aperture
  across all bands, and in the outer parts of a galaxy the colour changes
  only little with a further increase of the diameter.}, so we do not
expect any qualitative differences in the mean or median. The change in
sample size has only a negligible effect on the median values: we find
1.013 as compared to 1.011 for \jkoc\ and 0.310 as compared to 0.309 for
\hkoc .

\section{Galactic extinction}   \label{galext}

In Paper II we have used the colours of the bright sample to investigate
how well the extinction map used represents the line-of-sight Galactic
extinction. We can now do the same using the full sample. In the following
we distinguish between the already applied correction factor $f=0.86$ and
an additional factor \fa , which should be multiplied with $f$ to arrive at
a final value $f_{\rm new}$. From the slope of the extinction-corrected
\jko\ colour--extinction relation for the full sample, we have thus derived
\fa$=0.98\pm0.01$, which confirms that the correction factor derived in
Paper II is still acceptable.  For simplicity, the following discussion
will refer to \fa\ only.

In contrast to Paper II, we have excluded AGNs for this investigation
(about 100 galaxies). While the effect on \fa\ is minimal (0.982 versus
0.981 with AGNs included), the intercept becomes slightly redder by
$0\fm003$ (which if of the order of the error on the intercept) when AGNs
are included, emphasising their redder colour.

In Paper II, we mention the effect of applying an upper limit in $A_K$ on
the slope of the fitted regression line. We have reinvestigated this and
find a stable plateau for the the range $0\fm45 < {\rm lim}(A_K) \le 1\fm1$
for the new sample. Including galaxies at higher extinctions introduces a
bias due to on-sky variations in extinction smaller than the resolution of
the extinction maps. We thus use $A_K=1\fm1$ as the upper limit in our new
investigation. Otherwise, we use the same selection criteria as in Paper II
to ensure a photometrically clean sample.

\begin{figure} 
\centering \includegraphics[width=0.4\textwidth]{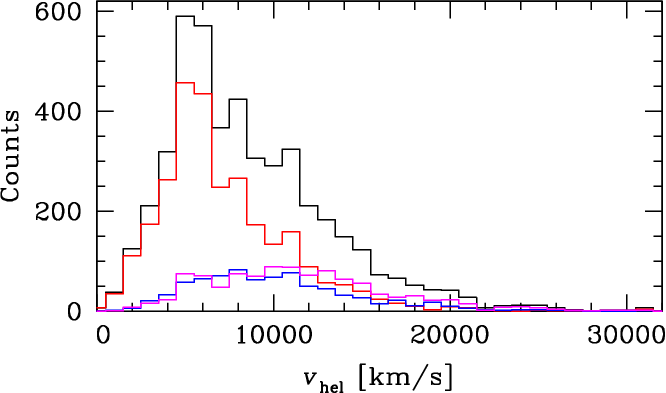}
\caption{Histogram of the redshift distribution of galaxies in the clean
  sample (black). The x-axis is truncated: there are six more galaxies
  between $v=32 000$ \kms\ and 56\,360 \kms . Also shown are the histograms
  for the subsamples S1 (red), S2 (blue) and S3 (magenta).
}
\label{histczplot}
\end{figure}

As discussed in Sect.~\ref{faint}, the fainter galaxies in our new sample
are often more distant, which may require a $k$-correction. We have done
two tests: firstly, we derived \fa\ for the samples S1, S2 and S3
separately, see Table~\ref{tabcolext}. We find that \fa\ slightly decreases
with the fainter samples, indicating that the $k$-correction indeed becomes
relevant. Figure~\ref{histczplot} shows the redshift distribution of the
various samples, confirming the shift towards higher redshifts for the
fainter samples. As a second test we have selected only galaxies that have
redshift information and compare how \fa\ changes when we apply the
$k$-correction: The latter gives \fa$=0.99\pm0.01$ as compared to
\fa$=0.98\pm0.01$ for the uncorrected $\sim4600$ galaxies with
redshift. The effect on the full sample is thus estimated negligible,
though we should keep in mind a possible selection effect where
predominantly brighter --- and thus less distant --- galaxies have been
measured for redshift. Nonetheless, we expect the effect to be small, that
is, less than the difference between the bright sample S1 and faint sample
S3 ($\Delta f < 0.02$). In addition, we note that the intercept
$1.00\pm0.00$ when using the $k$-correction is the same as the median
\jk\ colour for the high latitude bright sample (see
Table~\ref{tab2mhistb2}).

\begin{table}
\caption[]{Object samples at different stages during extraction.
\label{tabcolext}}
{\tiny
\begin{tabular}{lrrrr}
\hline
\noalign{\smallskip}
Sample &\multicolumn{1}{c}{\fa } & \multicolumn{1}{c}{intercept} & \multicolumn{1}{c}{$\sigma$} & \multicolumn{1}{c}{$N$} \\
\noalign{\smallskip}
\hline
\noalign{\smallskip}
Full sample        &  $0.982 \pm 0.007$  &  $1.034 \pm 0.003$  &  $0.12$  &  6155 \\ 
S1                 &  $0.990 \pm 0.009$  &  $1.026 \pm 0.003$  &  $0.11$  &  3178 \\ 
S2                 &  $0.980 \pm 0.015$  &  $1.037 \pm 0.006$  &  $0.11$  &  1189 \\ 
S3                 &  $0.970 \pm 0.014$  &  $1.045 \pm 0.005$  &  $0.12$  &  1788 \\ 
Galaxies with velocity  &  $0.978 \pm 0.009$  &  $1.033 \pm 0.003$  &  $0.10$  &  4604 \\ 
With $k$-correction   &  $0.991 \pm 0.009$  &  $0.996 \pm 0.003$  &  $0.10$  &  4604 \\ 
\noalign{\smallskip}
\hline
\end{tabular}
}
\end{table}

\begin{figure} 
\centering \includegraphics[width=0.5\textwidth]{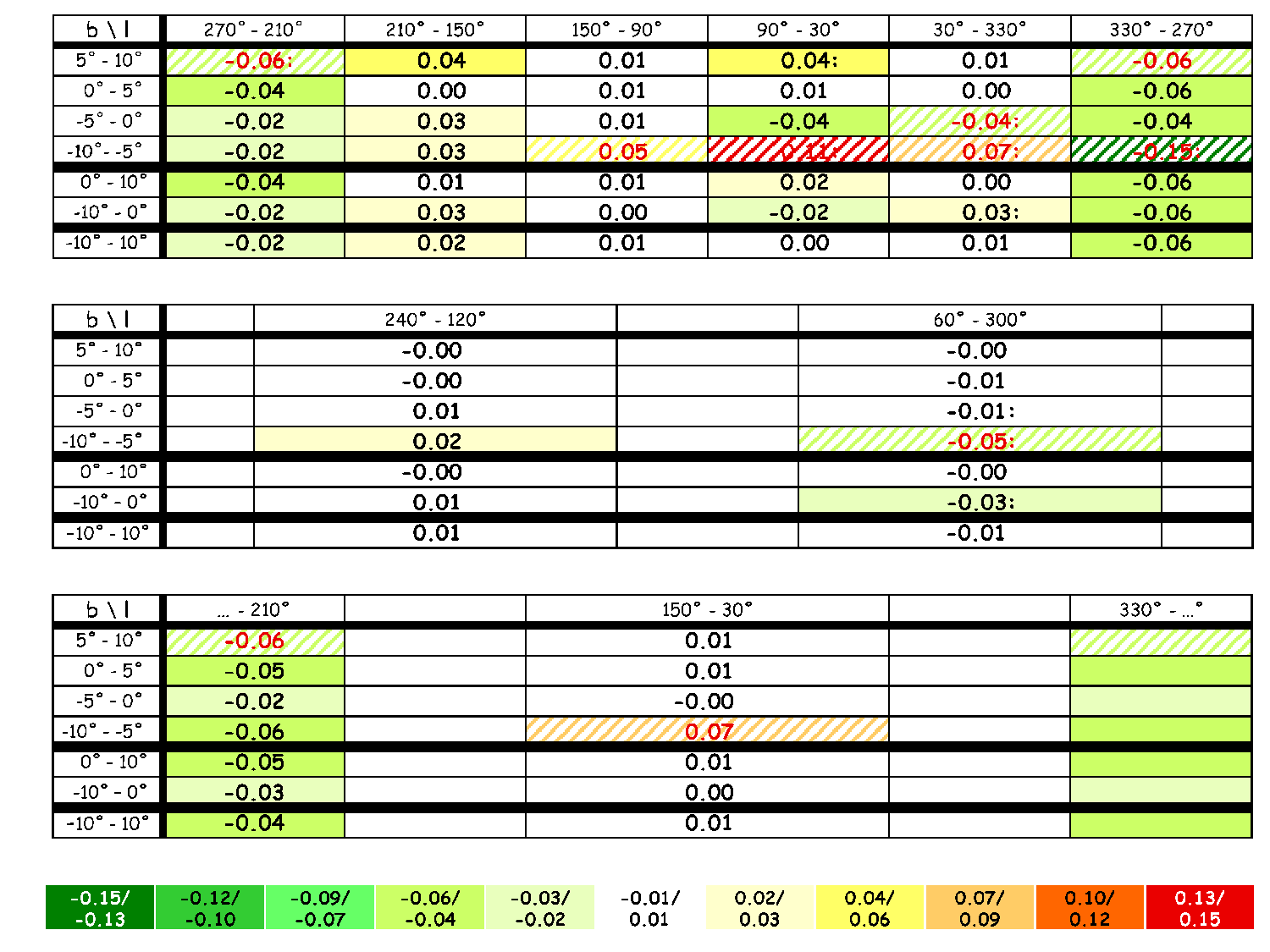}
\caption{Binned sky map with $\Delta f_{\rm a}$-values ($f_{\rm cell} -
  f_{\rm all}$) per cell. The colour scale is shown at the bottom. Red
  numbers may be overestimated, and colons indicate uncertain values. The
  $330\degr - 210\degr$ cells are wrapped around the Galactic Centre.
}
\label{plotglbmap}
\end{figure}

We can now derive an improved 'map' of the $f$-values using bins in
Galactic latitude and longitude. This will help to understand if there are
intrinsic variations in the extinction properties across the Galactic
plane. Since the binned samples will be much smaller than the full sample,
we keep the intercept fixed to the value derived for the full sample and
fit only the slope for each bin (see Sect.~5 in Paper II for details on the
derivation). Figure~\ref{plotglbmap} gives $\Delta f = f_{\rm cell} -
f_{\rm all}$, with the small, $60\degr \times 5\degr$ cells in the top four
rows, and larger, binned cells below (both in latitude, see, \eg rows
$5-7$, and in longitude, see the bottom two panels). Contrary to Paper II,
we use the neutral colour white for the range $-0.01 < \Delta f < 0.01$
(since 0.01 is the error on \fa\ in the unbinned sample). We do not apply
the $k$-correction so as to maximise the number of galaxies in the
individual bins and to avoid possible selection effects by uneven selection
of galaxies that have or have not redshift measurements. We do need to keep
in mind, though, a possible selection effect through large-scale structures
that may lead to more distant galaxies in one bin and not in another.

The map compares very well with the one for the bright sample (Fig. B4 in
Paper II) with the exception of a stronger deviation in the range $210\degr
< l < 330\degr$ which coincides with the direction of the dipole of the
cosmic microwave background (CMB). However, no deviation is seen in the
opposite direction. We need to keep in mind that overall the deviations
shown in the map are mostly smaller than $3 \sigma$ and thus not
significant. Nonetheless, the negative $\Delta f$-values seen for those
bins seem systematic and thus merit attention.

As a test, we have applied the $k$-correction, keeping in mind the above
mentioned caveats. The systematics in the range $210\degr < l < 270\degr$
have disappeared and lessened in the adjacent direction $270\degr < l <
330\degr$, indicating that the deviation we see in Fig.~\ref{plotglbmap} is
likely due to selection effects through cosmic variance (note that both the
Great Attractor as well as the Vela Supercluster are found in this
direction).

\section{Summary} \label{summary}

We have presented the full 2MZoA catalogue with an extinction corrected
magnitude limit of \ko$=11\fm75$, covering the ZoA at latitudes
$|b|<10\degr$ (ZOA sample) and areas of high extinction (\ebv $> 0\fm950$;
EBV sample) elsewhere. It thus serves as a complement to the 2MRS galaxy
catalogue, resulting in the uniquely all-sky e2MRS survey. The catalogue
comprises a total of 10\,493 objects, of which 6899 are galaxies. Of the
latter, 6757 are found in the ZOA sample, and 142 galaxies in the EBV
sample. Compared to the catalogue of bright (\ko\ $\le 11\fm25$) galaxies
presented in Paper I, we have elected to use a different map of foreground
extinction (the so-called GNILC extinction map by \citealt{planck16b}),
with a correction factor $f = 0.86$ (see Paper II). We have also updated
object classes (due to improved redshift and NIR information that have
become available in the meantime), added a flag to identify quasars and
active galaxies, revisited the extinction flag settings, and added redshift
information.

We have discussed the properties of this new catalogue and have
investigated possible differences to the bright sample (\ko\ $\le 11\fm25$)
due to the increased number of smaller and more distant galaxies. We have
found that the fainter sample is less contaminated by non-galaxian objects
like Galactic Nebulae. However, the fainter sample is more affected by
Galactic foreground extinction and is, in contrast to the bright sample,
not complete anymore even at low extinction levels. Stellar densities also
seem to have an increased effect on the completion levels. We also find
less redshift information in the literature for the fainter galaxies which,
often due to high foreground extinction, are more difficult to observe
spectroscopically. Based on the colours of the faintest galaxies ($11\fm50
< K^{\rm o}_{\rm s} \le 11\fm75$), many of these are distant galaxies and
would require a $k$-correction.

We have also compiled a supplementary sample of galaxies that become
brighter than the magnitude cut when we apply an additional
extinction-corrected diameter correction (see Paper I for details),
resulting in 1083 additional objects, of which 587 are galaxies. The
additional galaxies are predominantly found at high extinctions, improving
the completion levels of the combined (`extended') sample for extinctions
up to $\ak =1\fm3$ and possibly beyond.

Finally, according to the recipe presented in Paper II, we have compiled a
binned sky map of the correction factor $f$ to be applied to the Galactic
foreground extinction. Though we find larger deviations between cells than
in Paper II, they are all within the $3\sigma$ errors, and the possible
systematics found are likely due to cosmic variance. The full sample
results in a correction factor $f_{\rm new}=0.84\pm0.01$, but due to the
uncertainties in colours at the fainter end of our sample we continue to
recommend the correction factor $f = 0.86\pm0.01$ as derived in Paper II.

The next steps in completing this project are as follows. (1) Improving the
completion rate in the high stellar density area close to the Galactic
Bulge could be achieved via by-eye searches and/or machine learning
techniques (which might require additional by-eye verification, although
the number of objects that needed visual inspection would be greatly
reduced). Even more helpful would be searching the available deeper NIR
images where the star crowding is alleviated through the higher spatial
resolution of the images.  Several galaxy catalogues extracted from the VVV
and VVVX survey areas are already available (\eg \citealt{alonso25} and
references therein). The drawback is that combining these catalogues with
ours would lead to an inhomogeneously compiled 2MZoA catalogue which would
need careful calibration and adjustments.

(2) An improvement of galaxy photometry is important for the cosmic flow
field analysis. A semi-automated script that automatically subtracts stars
and extracts galaxy photometry combined with a thorough by-eye quality
control (\eg \citealt{said17}) would also remove some of the uncertainties
in the magnitude limit of the sample.

(3) For completing the redshift survey, we need to obtain the missing
redshifts. These would also help to improve the accuracy of the galaxy
photometry via a consistent application of the $k$-correction for all
galaxies. Redshift completeness can be achieved using the new generation of
radio telescopes, the SKA and SKA precursors, for \HI\ observations, and
large ground based telescopes like SALT and HET or space telescopes like
JWST for high quality IR of NIR spectroscopy.

Combined with the 2MRS, the here presented complementary catalogue forms
the extended 2MRS survey, e2MRS, which comprises a nearly complete and
homogeneous, magnitude-limited, all-sky galaxy catalogue. Apart from a
small area ($<4$\% of the whole sky) in the direction of the Galactic
Bulge, where galaxy detection is severely restricted by high stellar
densities, e2MRS can be considered as unbiased. As such it is invaluable
for studies of large-scale structures, flow fields and extinction across
the whole sky including the ZoA.

\section*{Data availability}

Tables A.1, A.2, A.3 and A.4 are available in electronic form at the CDS
via anonymous ftp to cdsarc.u-strasbg.fr (130.79.128.5) or via
http://cdsweb.u-strasbg.fr/cgi-bin/qcat?J/A+A/

\begin{acknowledgements}

The authors thank Lucas Macri for access to the most updated 2MRS data
set. This research has made use of data products from the Two Micron All
Sky Survey, which is a joint project of the University of Massachusetts and
the Infrared Processing and Analysis Center, funded by the National
Aeronautics and Space Administration and the National Science
Foundation. This research also has made use of the HyperLeda database, the
SIMBAD database, operated at CDS, Strasbourg, France, the NASA/IPAC
Extragalactic Database (NED) which is operated by the Jet Propulsion
Laboratory, California Institute of Technology, under contract with the
National Aeronautics and Space Administration and the Digitized Sky Surveys
which were produced at the Space Telescope Science Institute under
U.S. Government grant NAG W-2166. This research has made use of the VizieR
catalogue access tool, CDS, Strasbourg, France. The original description of
the VizieR service was published in \citet{vizier00}.  This research has
made use of the Astrophysics Data System, funded by NASA under Cooperative
Agreement 80NSSC25M7105.

\end{acknowledgements}




{\small
\bibliographystyle{aa} 
\bibliography{ZoA_bibfile_extended} 
}


\newpage
\begin{appendix}

\section{Tables} \label{apptab}

The four tables~\ref{tabzoa}\,--\,\ref{tabref} are available electronically
at the CDS. For clarity, we list in this section some example lines for
each table.

Table~\ref{tabzoa} presents the full catalogue. The columns are described
in Sect.~\ref{cat}.

For the radial velocity information in Table~\ref{tabzoa}, we searched the
literature for our objects.  We were mainly interested in original
measurements and do not list all available compilations. We searched online
databases (NED, Simbad, Leda) as well as catalogues available at the
CDS. Though we concentrated on objects classified as galaxies, we also
searched systematically for object class 5 (unknown) and 6 (lower
likelihood for being a galaxy) to be comprehensive.

The compilation of the velocities is given in Table~\ref{tabvop} for
optical velocities, and in Table~\ref{tabvhi} for \HI\ measurements.  The
tables give one line per velocity and reference per object. They are
ordered by (a) error (or quality) for original measurements, (b)
re-reductions (unless superseding the original publication), (c) by
publication year for referenced values (indicated by `c'; in Col.~2), and
(d) averaged values or non-detection entries. Non-accepted values (for
example detection of another galaxy nearby or wrong measurements) are
marked in the flag column.

The columns in the two tables are as follows: 
\begin{description}

\item{Col.\ 1:} ID: 2MASX catalogue identification number (based on J2000.0
  coordinates).

\item{Cols. 2 -- 5:} Flags repeated from Cols 9, 4, 5 and 6 from
  Table~\ref{tabzoa}.

\item{Col.\ 6:} Velocity measurement flags refering to the quoted
  publication. They are similar to the flags in Col. 10 of
  Table~\ref{tabzoa} with the addition of 'c' for compilation, `a' for
  compilation that gives averaged values, 'r' for re-reduced, and 'n' for
  non detections.

\item{Cols. 7 -- 9:} Redshift value and error in kilometre per second and
  the bibcode of the reference (see Table~\ref{tabref} for details).
  
\item{Col.\ 10:} Comments. An arrow indicates a quoted reference (usually
  also listed by us in this table), single quotes refer to comments made in
  the publication. Where possible we give telescope or instrument
  information (see also Table~\ref{tabref} for generic information
  regarding the publication).

\end{description}

Table~\ref{tabref} lists the references used in Tables~\ref{tabvop}
and~\ref{tabvhi} with: (1) Bibcode; (2) Name of first author; (3) Source of
data (either the wavelength regime or literature); (4) Observational and
velocity information where relevant (\eg\ telescope used, precision of the
velocity); (5) Comments.

\begin{table*}
\caption[]{2MZoA sample Part I (example lines). \label{tabzoa}}
{\tiny
\begin{tabular}{lrrrccccccccccccc}
\hline \hline
\noalign{\smallskip}
2MASX J          &  $l$ &  $b$ & EBV & \multicolumn{3}{c}{Object} & \multicolumn{2}{c}{Sample} & Disc  & 2MRS & \multicolumn{3}{c}{Vel flag} & NIR  & Phot & Ext  \\
                 &  deg &  deg & mag & cl    & off & obj          &  g & sobnei                & type  &      & Op    & HI & Tot             &      &        &    \\
(1)              & (2a) & (2b) & (3) & (4)   & (5) & (6)          &  \multicolumn{2}{c}{(7)}   & (8)   & (9)  & \multicolumn{3}{c}{(10)}     & (11) & (12)   & (13) \\
\noalign{\smallskip}
\hline
\noalign{\smallskip}
00000637+5319136 & 115.217 &  -8.780 &  0.303 & 1 & - &  - & g & 111111 & D4 &  c &    o &    - &    o & H &  - &  - \\
00093966+5301008 & 116.587 &  -9.339 &  0.390 & 1 & - &  - & g & 100110 & D2 & cp &    - &    - &    - & H &  P & e? \\
01475111+6305278 & 129.264 &   0.915 &  1.452 & 9 & e &  - & - & 111111 &  - &  - &    - &    - &    - & - & OC &  - \\
03044401+5406473 & 141.761 &  -3.812 &  1.006 & 1 & - &  - & g & 111111 & D1 &  e &    o & [h?] &    o & U &  - &  - \\
03382146+6419229 & 139.827 &   7.075 &  0.811 & 1 & - &  - & g & 100110 & D1 &  c &    o &    - &    o & - &  - &  - \\
05434399+1637482 & 190.314 &  -6.764 &  0.397 & 1 & - &  - & g & 100110 & D1 &  c &    o &    h &    h & H &  - &  - \\
06102125+1642343 & 193.430 &  -1.196 &  2.358 & 7 & - &  - & - & 111111 &  - &  - &    - &    - &    - & U &  - &  - \\
06553807-0705580 & 219.730 &  -2.303 &  0.982 & 1 & - &  - & g & 100110 & D1 &  - &    - &   h: &   h: & U &  - &  - \\
07561963-4137420 & 256.656 &  -6.723 &  0.799 & 1 & - &  q & g & 111111 & D1 &  c &    o &    - &    o & - &  - & e? \\
08210735-4321103 & 260.583 &  -3.796 &  2.319 & 1 & - &  - & g & 110111 & D1 &  - &    - &    - &    - & V &  - &  e \\
\noalign{\smallskip}
\hline
\noalign{\bigskip}
\end{tabular}
}
\addtocounter{table}{-1}
\caption[]{2MZoA sample Part II (example lines). }
{\tiny
\begin{tabular}{rrrrrrrrrrrrrrrl}
\hline \hline
\noalign{\smallskip}
$K_{20}$ & $H$-$K$ & $J$-$K$ & vc  & $a$ & $b/a$ & st.d. & \ko & ($H$-$K)^o$ & ($J$-$K)^o$ &$\ak$& $a^{\rm d}$& \kd & \multicolumn{2}{c}{vel} & Reference \\ 
  mag    & mag     &  mag    &     & ''  &       &       & mag & mag         &  mag        & mag & $''$       & mag & \multicolumn{2}{c}{\kms} & \\
  (14)   & (15)    & (16)    & (17) & (18)& (19)  & (20)  & (21)& (22)        & (23)        & (24)& (25)       & (26)& \multicolumn{2}{c}{(27)} & (28) \\
\noalign{\smallskip}
\hline
\noalign{\smallskip}
11.196 &  0.42 &  1.15 &   1 &  35.0 & 0.62 & 3.59 &  11.102 &   0.38 &   1.02 &  0.09 &   35.6 &  11.09 & 12447 &  65 & 2012ApJS..199...26H  \\ 
11.635 &  0.37 &  1.23 &   1 &  32.4 & 0.68 & 3.46 &  11.514 &   0.31 &   1.06 &  0.12 &   33.2 &  11.50 &   --- &  -- & ---                  \\ 
11.382 &  0.60 &  1.54 &  -2 & 104.8 & 0.44 & 3.85 &  10.932 &   0.39 &   0.91 &  0.45 &  126.5 &  10.73 &   --- &  -- & ---                  \\ 
 8.750 &  0.44 &  1.46 &   1 & 120.6 & 0.76 & 3.59 &   8.438 &   0.29 &   1.02 &  0.31 &  128.8 &   8.41 &  2467 &  29 & 2015ApJS..218...10V  \\ 
11.920 &  0.54 &  1.54 &   1 &  24.8 & 0.48 & 3.56 &  11.669 &   0.42 &   1.19 &  0.25 &   26.5 &  11.63 & 11272 & 300 & 1999MNRAS.308..897L  \\ 
11.502 &  0.39 &  1.37 &   1 &  46.4 & 0.24 & 3.47 &  11.379 &   0.33 &   1.20 &  0.12 &   47.6 &  11.36 &  8888 &  12 & 2017A\&A...599A.104T  \\
11.794 &  1.50 &  3.23 &   1 &  18.0 & 0.84 & 3.70 &  11.064 &   1.15 &   2.22 &  0.73 &   23.0 &  10.91 &   --- &  -- & ---                  \\ 
11.955 &  0.39 &  1.35 &   1 &  23.0 & 0.44 & 3.69 &  11.651 &   0.24 &   0.92 &  0.30 &   25.0 &  11.60 &  5571 &  20 & 2005A\&A...430..373T  \\
10.247 &  0.53 &  1.44 &   1 &  42.6 & 0.72 & 3.69 &   9.999 &   0.41 &   1.09 &  0.25 &   44.8 &   9.98 &  6558 &  17 & 2022ApJS..261....6K  \\ 
12.016 &  0.72 &  1.87 &   1 &  30.4 & 0.78 & 3.86 &  11.298 &   0.38 &   0.87 &  0.72 &   40.9 &  11.04 &   --- &  -- & ---                  \\ 
\noalign{\smallskip}
\hline
\noalign{\smallskip}
\end{tabular}
}
\end{table*}

\begin{table*}
\caption[]{Optical velocity information (example lines).
\label{tabvop}}
{\tiny
\begin{tabular}{lcccclrrll}
\hline \hline
\noalign{\smallskip}
2MASXJ ID  &  2MRS & \multicolumn{3}{c}{Object} & fl$_{\rm vel}$ & Vel & err & Reference & Comment \\
           &       & cl & off & obj &                 &     &           &         \\
\noalign{\smallskip}
\hline
\noalign{\smallskip}
00000637+5319136   & c  &   1 &  - &  -   &   o   &      12447 &  65 & 2012ApJS..199...26H  &   L \\
01345288+5524083   & c  &   1 &  - &  -   &   o   &       5765 &  50 & 1999PASP..111..438F  &   ('unpublished FAST observation'); H(err=5) \\
01345288+5524083   & c  &   1 &  - &  -   &   o?  &     [ 7643 &  22 & 1999ApJS..121..287H] &   (wrong match in NED, see \\
                   &    &     &    &      &       &            &     &                      &   \,\,\, 2MASXJ01343978+5525029) \\
04472448+4854366   & -  &   1 &  - &  -   &   o   &       3031 &  50 & 1994A\&AS..104..529T  &    \\
04472448+4854366   & -  &   1 &  - &  -   &   c   &       3031 &     & 2000MNRAS.317...55S  &   -> T94 \\
08294860-2208039   & c  &   1 &  - &  -   &   o:  &       6350 & 100 & 1994MNRAS.270...93Y  &   ('uncertain'; NED: err=20); H \\
08294860-2208039   & c  &   1 &  - &  -   &   c   &       6350 &     & 1993ApJS...89...57Y  &   -> Y94  \\
08294860-2208039   & c  &   1 &  - &  -   &   c   &       6350 &     & 1996ApJS..107..521V  &   -> Y94 \\
08294860-2208039   & c  &   1 &  - &  -   &   c   &       6350 &     & 2000MNRAS.317...55S  &   -> Y94 \\
21273973+4528586   & e  &   1 &  - &  -   &   o   &       5265 &     & 2001AJ....122.2381M  &  (Kitt Peak) \\
21273973+4528586   & e  &   1 &  - &  -   &   c   &       5265 &     & 2001AJ....122.2396D  &  -> M01 \\
\noalign{\smallskip}
\hline
\noalign{\smallskip}
\end{tabular}
}
\end{table*}

\begin{table*}
\caption[]{\HI\ velocity information (example lines).
\label{tabvhi}}
{\tiny
\begin{tabular}{lcccclrrll}
\hline \hline
\noalign{\smallskip}
2MASXJ ID  &  2MRS & \multicolumn{3}{c}{Object} & fl$_{\rm vel}$ & Vel & err & Reference & Comment \\
           &       & cl & off & obj &                 &     &           &         \\
\noalign{\smallskip}
\hline
\noalign{\smallskip}
00265906+6049388  &  -   &  1 &  -  & - &     h    &      7101 &   8 & 2018MNRAS.481.1262K &   ('marg') \\
00384223+6017130  &  -   &  1 &  -  & - &     h    &      4354 &   5 & 2018MNRAS.481.1262K &    \\
00384223+6017130  &  -   &  1 &  -  & - &     h    &      4355 &     & 1992ApJS...78..365H &   (noisy profile) \\
00384223+6017130  &  -   &  1 &  -  & - &     h:   &      4300 &     & 1987ApJ...320L..99K &    \\
00384223+6017130  &  -   &  1 &  -  & - &     c    &      4355 &  22 & 1992BICDS..41...31H &   -> H92 \\
00414439+5309441  &  c   &  1 &  -  & - &     n    &    [ 9717 &     & 2021A\&A...646A.113D]&   ('not detected') \\
02550583+6624065  &  e   &  1 &  -  & - &     h    &      3485 &     & 2009AJ....138.1938C &   (GBT; err~6) \\
04395830+4920510  &  e   &  1 &  -  & - &     h    &     -9999 &   7 & 2018ZOATF.NRT.0000K & ('confused')  \\ 
14300299-5234205  &  c   &  1 &  o  & - &     h?   &      4513 &     & 2004MNRAS.350.1195M &   (err=9; possible cross-match) \\
14300299-5234205  &  c   &  1 &  o  & - &     r?   &      4508 &     & 2005MNRAS.361...34D &    \\
22420452+5108042  &  c   &  1 &  -  & - &     m:   &      8017 &  15 & 1994A\&A...286...17S &   ('marg', snr=2.0); HL \\
\noalign{\smallskip}
\hline
\noalign{\smallskip}
\end{tabular}
}
\end{table*}

\begin{table*}  
\centering
\caption{{References for the redshifts (examples). } \label{tabref}}
{\tiny
\begin{tabular}{lllll}
\hline\hline
\noalign{\smallskip}
Bibcode & First author & Source & Obs/vel information & Comments \\
\noalign{\smallskip}
\hline
\noalign{\smallskip}
1967AJ.....72..821P & Page                  & opt              & v\_corr                           &   \\
1980Ap.....16..353M & Markarian             & opt              & SAO; z 4-digit                   & not systematical (no coordinates given); = 1980Afz....16..353M  \\
1982ApJ...255..373G & Gregory               & opt              & v\_galactocentric                 & converted in Leda  \\
1991srcp.book.....F & Fairall               & lit              & mostly averaged                  &   \\
1992ApJS...83...29S & Strauss               & opt, lit         &                                  & references -5xx and -1 are observations, others see ZCAT  \\
1999MNRAS.308..897L & Lawrence              & opt, lit         & WHT                              & QDOT; reference IDs see 2000MNRAS.317...55S  \\
2004MNRAS.350.1195M & Meyer                 & HI               & PKS, vres=26                     & blind survey; HIPASS; err from NED; = 2006HIPAS.C...0000: \\
                    &                       &                  &                                  & \,\,\, on NED  \\
2018MNRAS.481.1262K & Kraan-Korteweg        & HI               & NRT, vres=18                     &   \\
2022ApJS..261....2K & Koss                  & opt, lit         & z 8-digit                        & BASS-DR2; -> DR1  \\
\noalign{\smallskip}
\hline
\end{tabular}
}
\end{table*}

\section{Sky maps } \label{appmap} 

\begin{figure*} 
\vspace{-3cm}
\centering
\includegraphics[height=1.0\textwidth,angle=270,origin=c]{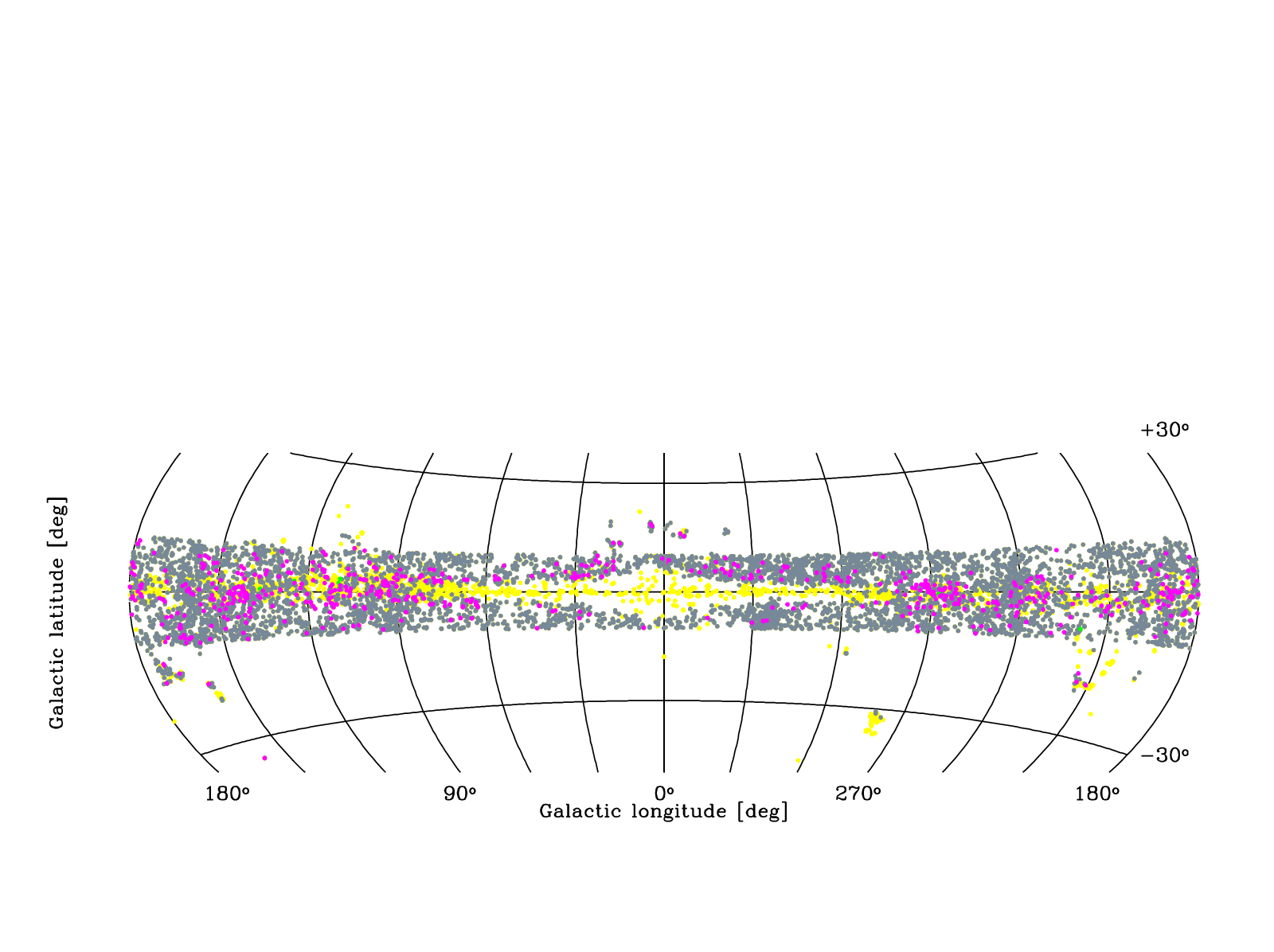}
\vspace{-3cm}
\caption{Aitoff projection in Galactic coordinates of the ZOA and EBV
  samples. Grey dots are galaxies, yellow dots non-galaxies, and green dots
  objects of unknown nature. Magenta dots indicate galaxies that are added
  through the diameter-depended extinction correction.
}
\label{mapplot}
\end{figure*}

The on-sky distribution of both the ZOA and the EBV samples is shown in
Fig.~\ref{mapplot}, with galaxies shown in grey and non-galaxies in yellow.
Magenta dots show galaxies from the extended sample (that is, with a
diameter-dependent extinction correction applied).  There are only a few
areas at higher latitudes ($b\geq10$\degr) that have extinctions high
enough to be in the EBV sample.

\end{appendix}


\end{document}